\documentclass[aps,preprintnumbers,floats, prd, twocolumn, nofootinbib]{revtex4-1}
\usepackage{graphicx}% Include figure files
\usepackage{dcolumn}% Align table columns on decimal point
\usepackage{bm}% bold math
\usepackage{amsmath}
\usepackage{amsthm}
\usepackage{multirow}
\usepackage{enumerate}
\usepackage{amsfonts}
\usepackage{ifthen}
\usepackage{psfrag}
\usepackage{slashed}
\usepackage{hyperref}
\usepackage{gensymb}
\usepackage[utf8]{inputenc}
\usepackage{color}
\usepackage{subfigure}
\usepackage{ulem}
\usepackage[utf8]{inputenc}
\usepackage{mathrsfs}
\usepackage{verbatim}
\usepackage{booktabs}
\hypersetup{colorlinks=true, citecolor=blue, linkcolor=blue, urlcolor=black}

\newcommand{\vev}[1]{\left\langle #1 \right\rangle}

\newcommand{\be}{\begin{equation}}
\newcommand{\ee}{\end{equation}}
\newcommand{\bea}{\begin{eqnarray}}
\newcommand{\eea}{\end{eqnarray}}

\begin{document}

\title{Probing Dark Matter with Future CMB Measurements}

\author{Junsong Cang$^{1,2}$}
\email{cangjs@ihep.ac.cn}
\author{Yu Gao$^{1}$}
\email{gaoyu@ihep.ac.cn}
\author{Yin-Zhe Ma $^{3,4}$}
\email{ma@ukzn.ac.za}

\affiliation{$^1$ Key Laboratory of Particle Astrophysics, Institute of High Energy Physics, Chinese Academy of Sciences, Beijing, 100049, China}
\affiliation{$^2$ School of Physical Sciences, University of Chinese Academy of Sciences, Beijing, 100049, China}
\affiliation{$^3$ School of Chemistry and Physics, University of KwaZulu-Natal, Westville Campus, Private Bag X54001, Durban, 4000, South Africa}
\affiliation{$^4$ NAOC–UKZN Computational Astrophysics Centre (NUCAC), University of KwaZulu-Natal, Durban, 4000, South Africa}

\begin{abstract}
The annihilation and decay of dark matter (DM) during the dark ages can leave imprints in the Cosmic Microwave Background Radiation (CMB) by altering the cosmic reionization process. CMB polarization anisotropy can be sensitive to such energy injection at higher redshift and therefore helps reducing degeneracy between spectral parameters in $\Lambda$CDM and other astrophysical parameters. In light of several upcoming CMB polarization experiments, such as AdvACTPol, AliCPT, CLASS, Simons Observatory, Simons Array, SPT-3G, we forecast their prospective sensitivity in probing dark matter annihilation and decay signals. We find that future missions have 95\% C.L. projected limits on DM decay and annihilation rates to orders of $\Gamma_\chi (\tau_{\chi}^{-1}) \sim 10^{-27}{~\rm{s}}^{-1}$ and $\left<\sigma v \right>/m_{\chi} \sim 10^{-29}{~\rm{cm^3s^{-1}GeV^{-1}}}$ respectively, significantly improving the sensitivity to DM from current experimental bounds.
\end{abstract}

\maketitle

\section{Introduction}
The majority of matter in the Universe exists in the form of non-luminous, non-baryonic Cold Dark Matter (CDM). A Weakly Interacting Massive Particle (WIMP) is a well-motivated particle physics candidate that explains the relic abundance of the Universe. Tremendous efforts have been invested for WIMP dark matter searches using terrestrial or space-borne experiments, see Refs.~\cite{Cushman:2013zza, Profumo:2013yn,Boveia:2018yeb,Schumann:2019eaa} for recent reviews. In addition to dedicated direct and indirect searches, the Cosmic Microwave Background (CMB) also provides an avenue for WIMP detection~\cite{Chen:2003gz,Adams:1998nr,Padmanabhan:2005es,Pierpaoli:2003rz}, which has gained increasing interest~\cite{Acharya:2019uba,Aghanim:2018eyx,Slatyer:2016qyl,Slatyer:2015jla,Liu:2016cnk,Kawasaki:2015peu}, especially with the availability of precision data from {\it Planck}~\cite{Aghanim:2019ame,Aghanim:2015xee} experiment. 

Thermally produced WIMP dark matter generally may annihilate or decay into the SM particles that subsequently become electromagnetically interacting electrons and photons. During the cosmic dark ages, this extra budget of energetic particle injection can heat up and ionize an amount of the neutral baryonic gas. Increased fraction of free electrons enhances the scattering of CMB photons during its propagation that leaves measurable imprints on the temperature and polarization anisotropy spectra of CMB. High-precision CMB measurement can place stringent limits on post-recombination annihilation and decay for dark matter over a wide mass range. The latest {\it Planck} 2018 results~\cite{Aghanim:2018eyx} report constraints on weak-scale DM annihilation cross-section that is comparable to the diffuse gamma-ray bound given by the combined analysis of Fermi-LAT and MAGIC data~\cite{Ahnen:2016qkx}. Besides, the relatively lower energy requirement for hydrogen ionization allows such limits to extend into much smaller mass ranges, effectively filling the gap between X-ray and gamma-ray indirect searches. 

Energy injection from dark matter modifies CMB anisotropy mostly via the ionization of intergalactic medium (IGM). In the temperature anisotropy spectrum, such injection is known to degenerate with several cosmological parameters~\cite{Padmanabhan:2005es} , such as the amplitude and tilt of primordial scalar perturbation ($A_{\rm{s}}$, $n_{\rm{s}}$), and the optical depth ($\tau$). Polarization anisotropy spectra help to break these degeneracies and gives tighter DM constraints. The {\it Planck} 2018 data~\cite{Aghanim:2019ame} provided the most stringent bounds on the cosmological parameters by far. Its constraints on DM annihilation cross-section by temperature anisotropy is tightened by about one order of magnitude with its polarization (EE) and cross-correlation (TE) spectra.

A number of operating or planned CMB experiments are expected to offer higher precision polarization data that may further enhance the sensitivity to dark matter annihilation and decay. These experiments include BICEP3/KECK Array \cite{Grayson:2016smb} and South Pole Telescope-3G \cite{Anderson:2018mry} in Antarctica, Advanced Atacama Cosmology Telescope Polarimeter (AdvACTPol) \cite{Calabrese:2014gwa,Li:2018uwb}, Cosmology Large Angular Scale Surveyor(CLASS) \cite{Essinger-Hileman:2014pja}, Simons Array \cite{Suzuki:2015zzg,Arnold:2014qym} and Simons Observatory \cite{Ade:2018sbj} in Chile, and Ali CMB Polarization Telescope (AliCPT) \cite{Li:2017lat} in China that aims at the northern hemisphere.

In this paper, we investigate the WIMP detection prospects of the upcoming CMB polarization experiments in a mass range of 10 KeV - 10 TeV, where the WIMPs are heavy enough so that their annihilation and decay products can efficiently ionize the neutral gas of the Universe. In the following Section~\ref{sec:inj} we briefly discuss the energy injection and deposition as well as the late-time clustering enhancement in annihilation rate. Section~\ref{sec:HyRecCAMB} discusses the impact of energy injection on the recombination history and CMB anisotropy and the breaking parameter degeneracies by polarization spectra. Section ~\ref{sec:Forecast} presents our forecasting method and analysis results, then we conclude in Section~\ref{sec:Discussion}.

\section{Energy injection and deposition}\label{sec:inj}
Assuming dark matter $\chi$ converts all its mass into the energy of annihilation and decay products, the energy injection rate per unit volume is
\be
\left(\frac{{\rm{d}}E}{{\rm{d}}V{\rm{d}}t}\right)_{\rm{INJ}}\ =
\left\{
\begin{array}{l}
m_{\chi} \Gamma_{\chi} n_{\chi} e^{-\Gamma_{\chi}t}  \ \ \ \ \ \ \ \ {\rm{decay}}\\
{g}\cdot 2m_{\chi} n_{\chi}^2 \left<\sigma v\right>   \ \ \ {\rm{annihilation}}
\end{array}
\right.
\label{eq:inja}
\ee
where $\Gamma_{\chi}$ is the DM decay width, defined as the inverse of decay lifetime ( $\Gamma_\chi=\tau_\chi^{-1}$). $g$ is a symmetry factor that equals 1/2 for self-conjugate dark matter, and 1/4 otherwise due to halving the number density between $\chi$ and $\bar{\chi}$. We will take $g=1/2$ throughout this paper. $ \left<\sigma v\right>$ is the thermally averaged annihilation cross section, here we only consider $s$-wave annihilation for which $\vev{\sigma v}$ is invariant. The average DM number density $n_\chi=\rho_{\rm{c}}\Omega_{\chi}(1+z)^3/m_{\chi}$, where $\Omega_{\chi}, \rho_{\rm c}$ are the cold dark matter fraction and the critical density of the Universe today.
As the relevant decay lifetime are much longer than the age of Universe, hereafter we will ignore the factor $e^{-\Gamma_{\chi}t}$ in Eq.(\ref{eq:inja}). The redshift dependent injection rates can be written as
\begin{eqnarray}
\left(\frac{{\rm{d}}E}{{\rm{d}}V{\rm{d}}t}\right)_{\rm{INJ}}^{\rm{dec}}\ &=&\  \Gamma_{\chi}\Omega_{\chi} (1+z)^3 \rho_{\rm{c}} \label{eq:dec} \\
\left(\frac{{\rm{d}}E}{{\rm{d}}V{\rm{d}}t}\right)_{\rm{INJ}}^{\rm{ann}}\ &=&\  \frac{\left<\sigma v\right>}{m_{\chi}}\Omega_{\chi}^2(1+z)^6\rho_{\rm{c}}^2 \label{eq:ann}
\end{eqnarray}
note that homogeneous DM distribution is assumed in Eq.(\ref{eq:ann}). The DM annihilation injection rate decreases faster as the Universe expands due to its dependence on higher power of $z$ . As will be discussed later, the annihilation rate will be boosted at late time by the DM clustering after the formation of halos. In comparison, DM decay yields a more steady rate of energy injection.

The final cascaded state of the annihilation and decay produces a variety of standard -model particles. Over the Hubble time-scale, these metastable products eventually decay into stable particles such as photons, electrons, protons and neutrinos. The injection of energy into the intergalactic medium is dominated by electrons and photons. Neutrinos do not interact efficiently with baryonic matter and decouple from the picture, protons are subdominant in abundance and can be ignored~\cite{Weniger:2013hja}.

The energetic photon and electrons lose their energy due to cosmic expansion, and also in a series of absorption and scattering processes with the CMB photons and the baryonic matter (mostly neutral gas). Interested readers can refer to~\cite{Kawasaki:2015peu,Slatyer:2009yq,Slatyer:2012yq,Belotsky:2014twa,Liu:2019bbm} for recent studies on the propagation and energy deposit of injected particles. Distortion in the CMB anisotropy is mostly due to increased ionization, and the most relevant energy deposit channels are
\begin{itemize}

\item Direct ionization of ground-state neutral hydrogen;

\item Excitation of the neutral hydrogen atom from $1s$ to $2p$ state, contributing to indirect ionization.
%\item Heating of the gas;
\end{itemize}

A handful of other energy loss channels are less effective in altering the anisotropy spectra. For example,  energy deposit into heating the IGM can cause a dramatic rise in the gas temperature at low redshift, yet the impact on anisotropy spectra is insignificant compared to ionization channels. Contribution from helium ionization is found to be subdominant compared to hydrogen~\cite{Liu:2016cnk} and will not be included in our calculation. A fraction of injected energy can also be deposited into changing the energy spectrum of CMB.

For highly relativistic injected particles, their energy deposition is a gradual process that continues to later times. For a given redshift, an effective efficiency $f_{\rm{c}}$ represents the ratio between the rate of energy deposition to that of DM injection at the same redshift,
\begin{equation}
\left(\frac{{\rm{d}}E}{{\rm{d}}V{\rm{d}}t}\right)_{\rm{DEP,c}}\ =\ f_{\rm{c}}\left(\frac{{\rm{d}}E}{{\rm{d}}V{\rm{d}}t}\right)_{\rm{INJ}}
\end{equation}
here the subscript c labels the deposition channel, DEP and INJ refer to deposition and injection rates respectively. The deposition efficiency $f_{\rm{c}}$ depends on the particle species, its energy upon injection, the redshift, and accumulates over particles injected at earlier times. A previously popular scheme uses an ``SSCK" prescription \cite{1985ApJ...298..268S,Chen:2003gz,Slatyer:2015jla} that assumes a fraction $(1-x_{\rm{e}})/3$ of the energy deposit goes into ionization. Here, we will adopt $f_{\rm{c}}$ from recent numerical analyses.

\subsection{Deposition efficiency}

The effective energy deposition efficiency $f_{\rm{c}}$ can be constructed from a discretized deposit fraction coefficient $T_{{\rm{c}},ijk}(z_i,E_j,z_k)$ over redshift and energy bins, as given in Ref.~\cite{Slatyer:2015kla}, which describes the 
fraction of $E_j$ deposited into channel $\rm{c}$ at redshift $z_i$, where $E_j$ is the particle's initial kinetic energy at its injection redshift $z_k$.
$f_{\rm{c}}(z_i)$ is obtained by summing $T$ over all injection redshift bins prior to $z_i$,
\begin{equation}
f_{\rm{c}}(z_{i})\approx \frac{\sum_{j}\sum_{k}A_{jk}{\rm d}V(z_{k}){\rm d}t(z_k)T_{{\rm c},ijk}
}{\sum_{j}A_{ji}{\rm d}V(z_{i}){\rm d}t(z_{i})}
\label{eq:fc}
\end{equation}
where 
\begin{eqnarray}
A_{jk}=E_j \left(\frac{{\rm d}N}{{
\rm d}E_j {\rm d}V(z_{k}){\rm d}t(z_{k})} \right) {\rm d}E_{j}
\end{eqnarray}
In Eq.~(\ref{eq:fc}), ${\rm{d}}t(z)$ is the time interval corresponding to ${\rm{d}}\ln(1+z)=10^{-3}$.
The numerator sums over the contribution from the earlier injection, and the denominator gives the total energy injection at $z_i$. From Eq.(\ref{eq:fc}) it is clear that $f_{\rm{c}}$ for annihilation and decay scenarios are differently weighted over historical deposits. Also, $f_{\rm{c}}$ may exceed unity at the late time due to accumulated injection.

We will consider direct production of $e^\pm$ and $\gamma\gamma$, for which the injection spectrum is monochromatic (at $E_j$),
\begin{equation}
\frac{{\rm{d}}N}{{\rm{d}}E{\rm{d}}V{\rm{d}}t}=\frac{1}{E_j}\left(\frac{{\rm{d}}E}{{\rm{d}}V{\rm{d}}t}\right)_{\rm{INJ}}\delta_{\rm{D}}(E-E_j)
\end{equation}
where $\delta_{\rm{D}}$ refers to Dirac delta function. Eq.(\ref{eq:fc}) can then be simplified as
\begin{equation}
f_{\rm{c}}(z_{i},E_j)=\frac{H(z_{i})}{(1+z_{i})^{\beta}}\sum_{k}\frac{T_{{\rm c},ijk}{(1+z_{k})^{\beta}}}{H(z_{k})}
\end{equation}
where $\beta=0$ for DM decay and $\beta=3$ for DM annihilation, note that we used $H(z)=-{\rm{d}}\ln(1+z)/{\rm{d}}t$.

\subsection{Clustering enhancement}

Structure formation at low $z$ let DM cluster into halos, and the DM density condensation enhances DM annihilation due to $\rho^2$ dependence. This enhancement over unclustered average-density annihilation is formulated as a `boost factor' $B(z)$~\cite{Liu:2016cnk},
\begin{equation}
\left(\frac{{\rm{d}}E}{{\rm{d}}V{\rm{d}}t}\right)^{\rm{ann, boosted}}_{\rm{INJ}}\ =\ \left[1+B(z)\right] \ \left(\frac{{\rm{d}}E}{{\rm{d}}V{\rm{d}}t}\right)^{\rm{ann}}_{\rm{INJ}}
\label{eq:boost}
\end{equation}
The overall boost is obtained by integrating over the contribution from halos~\cite{Taylor:2002zd},
\be
B(z)=\frac{\Delta_{\rm c} \rho_{\rm c}}{\rho^2_{\rm{DM}}}\int^{\infty}_{M_{\rm min}}MB_{\rm h}(M)\frac{{\rm{d}}n}{{\rm{d}}M}{\rm{d}}M
\ee
where $\Delta_{\rm c} \rho_{\rm c}$ is the average density of bound halos, in subsequent analysis we will assume $\Delta_{\rm c}=200$ and use a cut-off of $M_{\rm{min}}=10^{-6}\rm{M}_{\odot}$ as a reasonable estimate for the minimum halo mass. $B_{\rm{h}}(M)$ is the enhancement from an individual halo of mass $M$,
\be
B_{\rm h}(M)=\frac{4\pi}{\bar{\rho}_{\rm h}^2V_{\rm h}(M)}\int^{r_{200}}_0{\rm{d}}r\rho^2(r)r^2
\ee
here $\bar{\rho}_{\rm h}$ gives the average density of the halo distributed within volume $V_{\rm h}(M)$. $\rho(r)$ describes the radial density profile, truncated at a virial radius $r_{200}$.  
We consider spherical collapse model of halo formation, for which the mass function is given by Ref.~\cite{Schneider:2013ria,Sheth:1999mn},
\bea
\frac{{\rm{d}}n}{{\rm{d}}\ln M}&=&\frac{1}{2}f({\rm{\nu}})\frac{\rho_{\rm{DM}}}{M}\frac{{\rm{d}}\ln({\rm{\nu}})}{{\rm{d}}\ln M}\\
f({\rm{\nu}})&=&A\sqrt{\frac{2q{\rm{\nu}}}{\pi}}[1+(q{\rm{\nu}})^{-p}]\ {\rm{e}}^{-q{\rm{\nu}}/2}
\eea
with $A=0.3222$, $p=0.3$, $q=0.707$ and the scaled variable 
$\nu \equiv [\delta_{\rm{cr}}/(\sigma(M) D(z) )]^2 $, where $\delta_{\rm{cr}} \approx 1.686$, $\sigma(M)$ is the rms linear over-density and $D(z)$ is the growth factor normalized to unity today.

 We ignored further boost from halo substructures and used a typical cuspy Einasto profile~\cite{1965TrAlm...5...87E} for the main halo,
\begin{equation}
\rho(r)=\rho_{-2}\ {\rm exp}\left({-\frac{2}{\alpha_{\rm{e}}}\left[\left(\frac{r}{r_{-2}}\right)^{\alpha_{\rm{e}}}-1\right]}\right)
\end{equation}
where $\rho_{-2}$, $r_{-2}$, $\alpha_{\rm{e}}$ are halo profile parameters, assumed to follow the empirical relations given in Ref.~\cite{Klypin:2014kpa}, $\alpha_{\rm e}=0.115+0.0165\mu^2$, and $r_{200}/r_{-2}=6.5\mu^{-1.6}(1+0.21\mu^2)$, where $\mu \equiv \delta_{\rm{cr}}/\sigma(M)$. $B(z)$ reaches unity around $z=45$ and increases dramatically afterwards, giving $B=783$ at $z=20$. Having acquired the boost factor, we can account for the clustering enhancement by using a `boosted' version of the deposition efficiency for annihilation,
\begin{equation}
f_{\rm{c}}^{\rm{boost}}(z_{i},E_j)=\frac{H(z_{i})}{(1+z_{i})^3}\sum_{k}\frac{T_{{\rm c},ijk}{(1+z_{k})^3}(1+B(z_{k}))}{H(z_{k})}
\end{equation}

%-------------------------------------------------------------- FIG: HyRec
\begin{figure}[t]
\centering
\includegraphics[width=9cm]{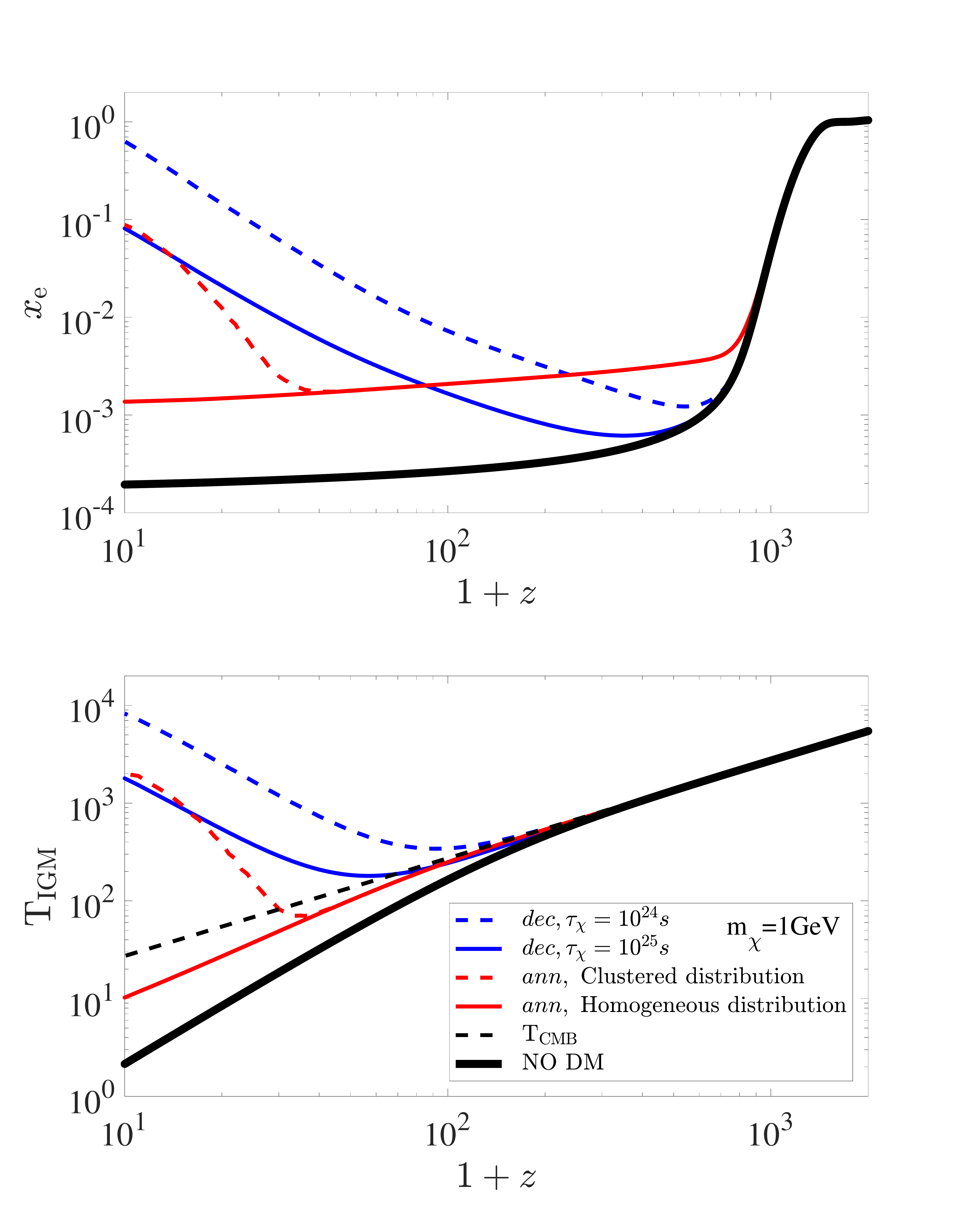}
\caption{Illustrative ionization fraction (top) and intergalactic medium temperature (bottom) evolution for DM annihilation (ann) and decay (dec) into $e^-e^+$, assuming DM mass $m_{\chi}=1\,{\rm{GeV}}$.
Both annihilation (red) curves assume $\left< \sigma v \right> =3\times 10^{-26}\,{\rm cm^{3}}\,{\rm s}^{-1}$ where the dashed/solid linetypes denote the annihilation scenario with/without late-time halo boost. For DM decay (blue) the dashed (solid) curve represents a decay lifetime at $10^{24}$( $10^{25}$) seconds respectively. The standard evolution of ionization fraction and IGM temperature without DM injection are shown in solid black curves. CMB temperature is shown in the lower panel  (black dashed) for comparison. The legend applies to all panels.}
\label{fig:HyRec}
\end{figure}

%-------------------------------------------------------------- FIG: CAMB
\begin{figure}[t]
\centering
\includegraphics[width=9cm]{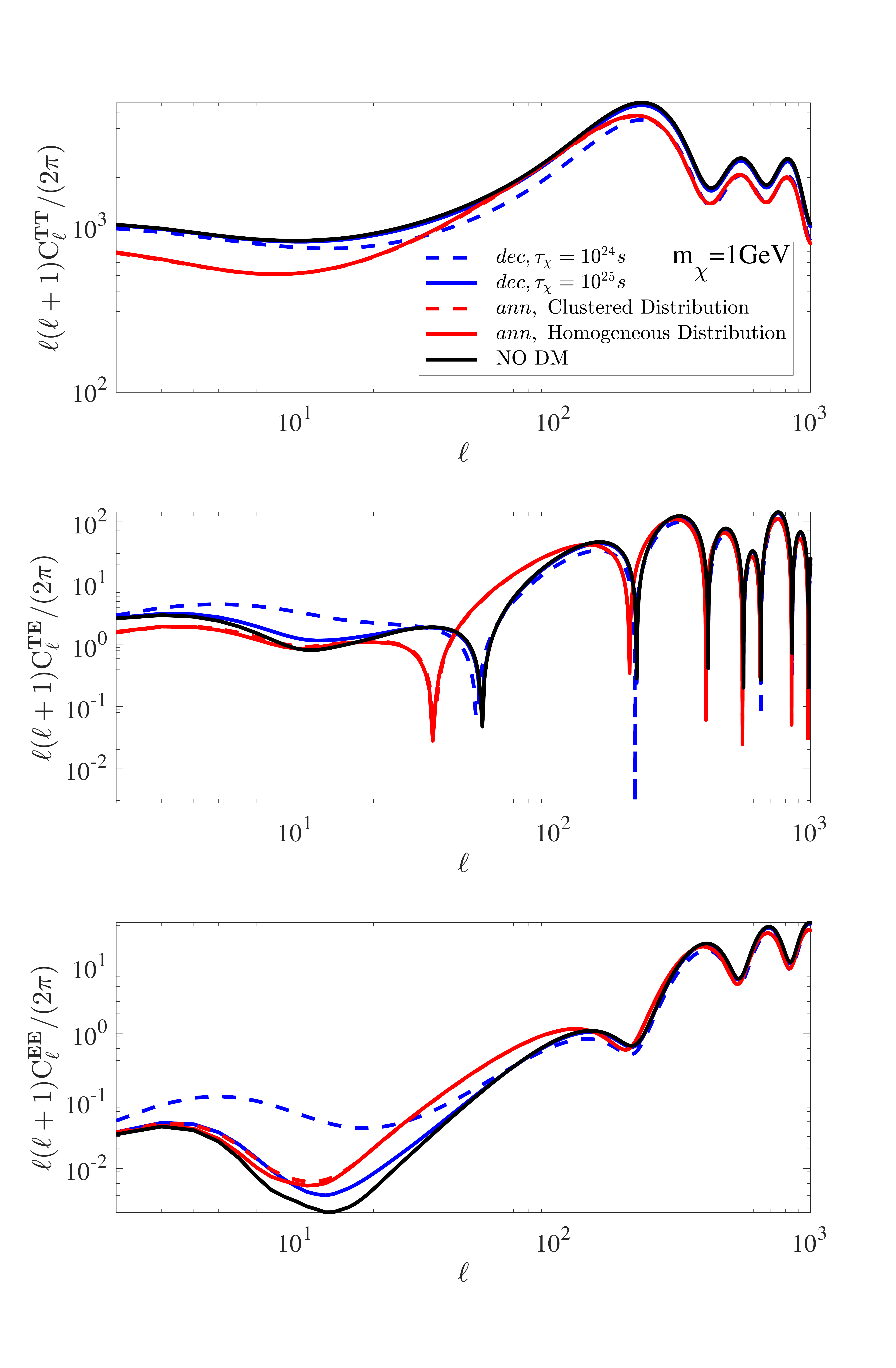}
\caption{
CMB TT (top), TE (middle), EE (bottom) anisotropy spectra for the same injection scenarios as in Fig.~\ref{fig:HyRec}.}
\label{fig:CAMB}
\end{figure}

\section{Ionization history} \label{sec:HyRecCAMB}
DM induced energy deposit can be incorporated into the ionization and IGM temperature evolution equations as~\cite{Chen:2003gz,Slatyer:2016qyl}
\bea
\frac{{\rm{d}}x_{\rm{e}}}{{\rm{d}}z}&=&\left(\frac{{\rm{d}}x_{\rm{e}}}{{\rm{d}}z}\right)_{\rm{0}}-\frac{I_{\chi}}{(1+z)H(z)}
\label{eq:dxdz}
\eea
\begin{flalign}
\frac{{\rm{d}}T_{\rm{IGM}}}{{\rm{d}}z}&=&\left(\frac{{\rm{d}}T_{\rm{IGM}}}{{\rm{d}}z}\right)_{\rm{0}} -\frac{2}{3k_{\rm B}(1+z)H(z)}\frac{K_{\rm{h}}}{1+f_{\rm{He}}+x_{\rm{e}}}
\label{eq:dtdz}
\end{flalign}
where $x_{\rm{e}}$ and $T_{\rm{IGM}}$ are the ionization fraction and IGM temperature. $k_{\rm{B}}$ is the Boltzmann constant and $f_{\rm{He}}$ is the helium fraction. The terms with subscript 0 represents the unaltered  evolution equations in standard cosmology~\cite{AliHaimoud:2010dx,Liu:2016cnk}. $I_{\chi}$ and $K_{\rm{h}}$ are the ionization and heating terms introduced by DM. Enhancement to the IGM temperature does not directly contribute to altering the anisotropy spectra, 
but it is of interest for the epoch of reionization observations. 
The ionization term $I_{\chi}$ can be further decomposed into hydrogen ionization from ground state ($I_{\chi i}$) and from $n=2$ state ($I_{\chi \alpha}$),
\bea
I_{\chi i}(z)&=&\frac{1}{n_{\rm{H}}(z)E_i}\left(\frac{{\rm{d}}E}{{\rm{d}}V{\rm{d}}t}\right)_{{\rm{DEP}},i} \\
I_{\chi \alpha}(z)&=&\frac{1-C}{n_{\rm{H}}(z)E_\alpha}\left(\frac{{\rm{d}}E}{{\rm{d}}V{\rm{d}}t}\right)_{{\rm{DEP}},\alpha}
\eea
and the IGM heating term $K_{\rm{h}}$ reads
\begin{equation}
K_{\rm h}(z)=\frac{1}{n_{\rm{H}}(z)}\left(\frac{{\rm{d}}E}{{\rm{d}}V{\rm{d}}t}\right)_{{\rm{DEP}},h}
\end{equation}
where the subscripts $i, \alpha, h$ denote the corresponding deposition channels that was indicated by the subscript $\rm{c}$ in previous sections. $n_{\rm{H}}$ is the number density of hydrogen nuclei, $E_i=13.6$  eV is the ionization energy for ground state hydrogen atom, and $E_{\alpha}=10.2$ eV is the difference in binding energies between the 1$s$ and 2$p$ states. $C$ is the probability for an $n=2$ state hydrogen atom to transit back to $n=1$ state before it gets ionized ~\cite{Chen:2003gz,Slatyer:2016qyl}.

% -------------------------------------------------------- FIG:DEGENERACY
\begin{figure*}[htp]
\centering
\subfigbottomskip=-200pt
\subfigcapskip=-7pt
\includegraphics[width=16cm]{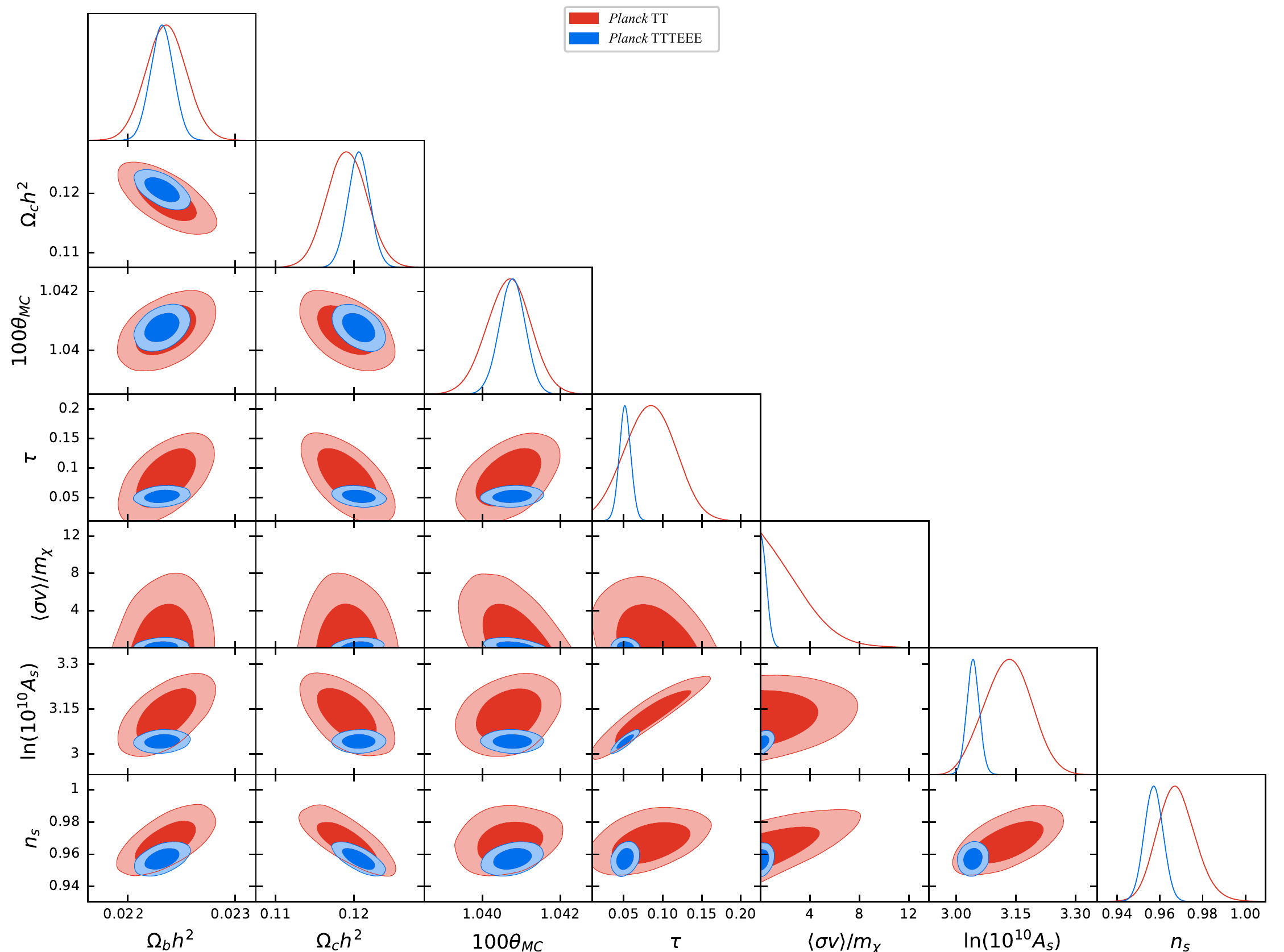}
\caption{
Marginalized constraints on DM injection ( parameterized by $\left<\sigma v\right>/m_{\chi}$ in $10^{-27} \rm{cm^3\,s^{-1}\,GeV^{-1}} $) and base $\Lambda$CDM parameters, set by {\it{Planck}} 2018 TT ( red ) and TT+TE+EE ( blue ) data. 
Dark and light contours in off-diagonal figures indicate 95\% and 68\% CL bounds respectively. DM mass is set to 100\,GeV and assumed to annihilate into $e^-e^+$ under homogeneous distribution scenario.
}
\label{fig:DEGENERACY_Triangular_Plot}
\end{figure*}

We used modified HyRec \cite{AliHaimoud:2010dx} and the CAMB~\cite{Lewis:1999bs} codes to calculate the ionization and temperature history before reionization and the CMB anisotropy spectra. 
Fig.~\ref{fig:HyRec} shows the ionization fraction and IGM temperature evolution for a few examples of annihilation and decay scenarios. It shows both DM annihilation scenarios at the thermal relic's typical cross-section, and DM decay at the currently allowed rate $10^{-25}$ s$^{-1}$ can lead to an order of magnitude increase in the ionization fraction and IGM temperature.
The corresponding anisotropy $C_{\ell}$ spectra are shown in Fig.~\ref{fig:CAMB}.

Energy deposits from DM will increase the ionization fraction and IGM temperature after recombination, this broadens the surface of the last scattering and suppresses CMB temperature correlation on scales smaller than the width of the surface~\cite{Adams:1998nr,Padmanabhan:2005es}.  Unfortunately, this effect degenerates with several cosmological parameters, including the spectral index ($n_{\rm{s}}$) and amplitude ($A_{\rm{s}}$) of the primordial scalar perturbation, and the optical depth ($\tau$). 
However, as shown in Fig.\ref{fig:CAMB}, energy injection near recombination can cause visible shifts in polarization anisotropy peaks while enhancing EE correlations at large scales~\cite{Padmanabhan:2005es}. This feature allows polarization data to break the degeneracy with the spectral amplitude and tilt. 
In addition, this high-$z$ injection dependence significantly reduces degeneracy with the overwhelming astrophysical injection at reionization epoch.

Fig.\ref{fig:DEGENERACY_Triangular_Plot} visualizes the correlation between DM annihilation injection (parameterized by $\left<\sigma v\right>/m_\chi$) and the base $\Lambda$CDM cosmological parameters. The constraints are obtained by fitting {\it Planck} 2018 TT (red) and TT+TE+EE (blue) data with the CosmoMC codes~\cite{Lewis:2002ah,Lewis:2013hha}. Tilting in the elliptical contour's axis indicate parameter correlation. The resulting linear correlation coefficients between $\left<\sigma v\right>/m_\chi$ and six relevant cosmological parameters are listed in Tab. I. With the inclusion of polarization we observe a significant reduction of parameter degeneracies.

DM clustering boost has a huge impact on $x_{\rm{e}}$ and $T_{\rm IGM}$ at late time. However, because CMB is rather insensitive to energy injection at low redshift, the changes to the anisotropy spectra are relatively minor, as demonstrated most evidently in TT and EE panels in Fig.~\ref{fig:CAMB}.

\begin{table}
\begin{tabular}{c|cc}
\hline \hline
 &\ \  TT\ \  & TT,TE,EE \\
 \hline
$\tau$ &-0.22 &-0.04 \\
$n_{\rm s}$ &0.59 &0.14 \\
$\ln(10^{10}A_{\rm s})$ &0.17 &0.27 \\
$\Omega_{\rm b} h^2$ &0.17 &0.07 \\
$\Omega_{\rm c} h^2$ &-0.08 &0.16 \\
$100\theta_{\rm MC}$ &-0.37 &-0.28 \\
\hline \hline
\end{tabular}
\label{tab:Linear_Corre}
\caption{Linear correlation coefficients between $\left<\sigma v\right>/m_\chi$ and cosmological parameters corresponding to Fig.\ref{fig:DEGENERACY_Triangular_Plot}.}
\end{table}

\section{Sensitivity Estimates} 
\label{sec:Forecast}

Several upcoming observations are expected to conduct high precision measurement of CMB polarization anisotropy at $\mu {\rm{K}}$-${\rm{arcmin}}$ sensitivities with arcmin beams, including AdvACTPol~\cite{Calabrese:2014gwa,Li:2018uwb}, AliCPT~\cite{Li:2018rwc,Li:2017lat}, CLASS~\cite{Essinger-Hileman:2014pja}, LiteBird~\cite{Hazumi:2019lys}, Simons Array~\cite{Suzuki:2015zzg,Arnold:2014qym}, Simons Observatory~\cite{Ade:2018sbj}, SPT-3G ~\cite{Anderson:2018mry}, and many more proposed for the future~\cite{Abazajian:2016yjj, Hanany:2019wrm, Calabrese:2016eii, DiValentino:2016foa}. We will study the expected sensitivities on DM injection of a few missions that are either already operational or will be so in near future.

\subsection{Forecasting Procedure}

Assuming null signal, a sensitivity bound on DM energy injection can be placed by testing the spectral deviation from fiducial TT, TE and EE anisotropy spectra $\hat{C_{\ell}}$ that serve as fake data. We use the ``exact full sky" likelihood function \cite{Hamimeche:2008ai} for significance calculation,
\bea
-2{\ln}\mathcal{L}(\{C_{\ell}\}|\{{\hat{C}_{\ell}}\})=\ \ \ \ \ \ \ \ \ \ \ \ &&   \\ \label{eq:likelihood}
 f_{\rm sky}\times \sum_{\ell}(2\ell +1)\{{\rm{Tr}}[{\hat{C}_{\ell}}C_{\ell}^{-1}]& -&{\ln}|{\hat{C}_{\ell}}C_{\ell}^{-1}|-2\} \nonumber
\eea
where $f_{\rm{sky}}$ is the fraction of sky covered by an experiment, $C_{\ell}$ is a function of cosmological and DM parameters, while $\hat{C_{\ell}}$ is a simulated anisotropy spectra serving as mocked data. With both temperature (T) and polarization (E) correlations, $C_{\ell}$ and $\hat{C_{\ell}}$ are $2 \times 2$ matrices:
\begin{equation}
\label{eq:covariance_definition} 
C_{\ell} \equiv 
\begin{bmatrix}
C_{\ell}^{TT} & C_{\ell}^{TE} \\
C_{\ell}^{TE} & C_{\ell}^{EE}\\
\end{bmatrix}
\end{equation}

\begin{equation}
\label{eq:covariance_definition} 
\hat{C}_{\ell} \equiv 
\begin{bmatrix}
\bar{C}_{\ell}^{TT} + N_{\ell}^{TT} & \bar{C}_{\ell}^{TE} \\
\bar{C}_{\ell}^{TE} & \bar{C}_{\ell}^{EE} + N_{\ell}^{EE}\\
\end{bmatrix}
\end{equation}

The fixed `fiducial' $\bar{C}_{\ell}$ is generated using {\it Planck} 2018 best-fit ~\cite{Aghanim:2018eyx} $\Lambda$CDM parameters  ($\Omega_{\rm{b}}h^2=0.0224$, $\Omega_{\rm{c}}h^2=0.1193$, $100\theta_{{\rm{MC}}}=1.041$, $\tau=0.056$, $\ln(10^{10}A_{\rm{s}})=3.047$, $n_{\rm{s}}=0.967$), without DM injection. $N_{\ell}^{TT}$ and $N_{\ell}^{EE}$ are the instrumental white noise power spectra, for a multi-frequency CMB experiment they are given by \cite{Knox:1995dq,Errard:2015cxa}:
\bea
N^{{\rm{EE}}}_{\ell}&=&\left[\sum_{\nu} \omega_{E,\nu}{\exp}\left(-\ell (\ell +1)\frac{\theta^2_{\rm{FWHM},\nu}}{8\ {\ln}2}\right)\right]^{-1} \\
\label{eq:Nlv}
N^{{\rm{TT}}}_{\ell} &=&\frac{1}{2}\ N^{{\rm{EE}}}_{\ell}
\eea
here the subscript $\nu$ labels the frequency channel, $\omega_{{\rm{E}},\nu}^{-\frac{1}{2}}$ is the white noise level in $\mu {\rm{k}}\cdot  {\rm{rad}}$. $\theta_{\rm{FWHM},\nu}$ denotes the full width at half maximum beam size in radians.  Specifications for experiments considered are collected in Appendix \ref{apd:Specs}.

In addition to detector noises, residual foreground would be a contamination that contributes to $\hat{C}_{\ell}$. However it is beyond the scope of this paper to make robust foreground removal estimate for each experiment, and  we present the results assuming the anisotropy foreground has been successfully subtracted.

\subsection{Prospective limits}

Here we present the prospective limits of DM annihilation and decay rates. Both $m_{\chi}^{-1} \left<\sigma v\right>$ and $\Gamma_{\chi}$ are considered time and velocity independent. The dark matter parameters $m_\chi$, $m_{\chi}^{-1} \left<\sigma v\right>$ and $\Gamma_{\chi}$ are included into CosmoMC as new variables.  We marginalize over $\Lambda$CDM parameters to derive the bounds on dark matter papers. Experimental nuisance parameters are also marginalized over in case {\it{Planck}} data are included. CosmoMC uses a preset Gelman-Rubin ``mean of chain variance" $R$ value as the convergence criterion in the Markov chain process, and we ensure $R-1 \le$ 0.01 in the results. 

\begin{figure*}[htp]
\centering
\subfigbottomskip=-200pt
\subfigcapskip=-7pt
\subfigure{\includegraphics[width=9.5cm]{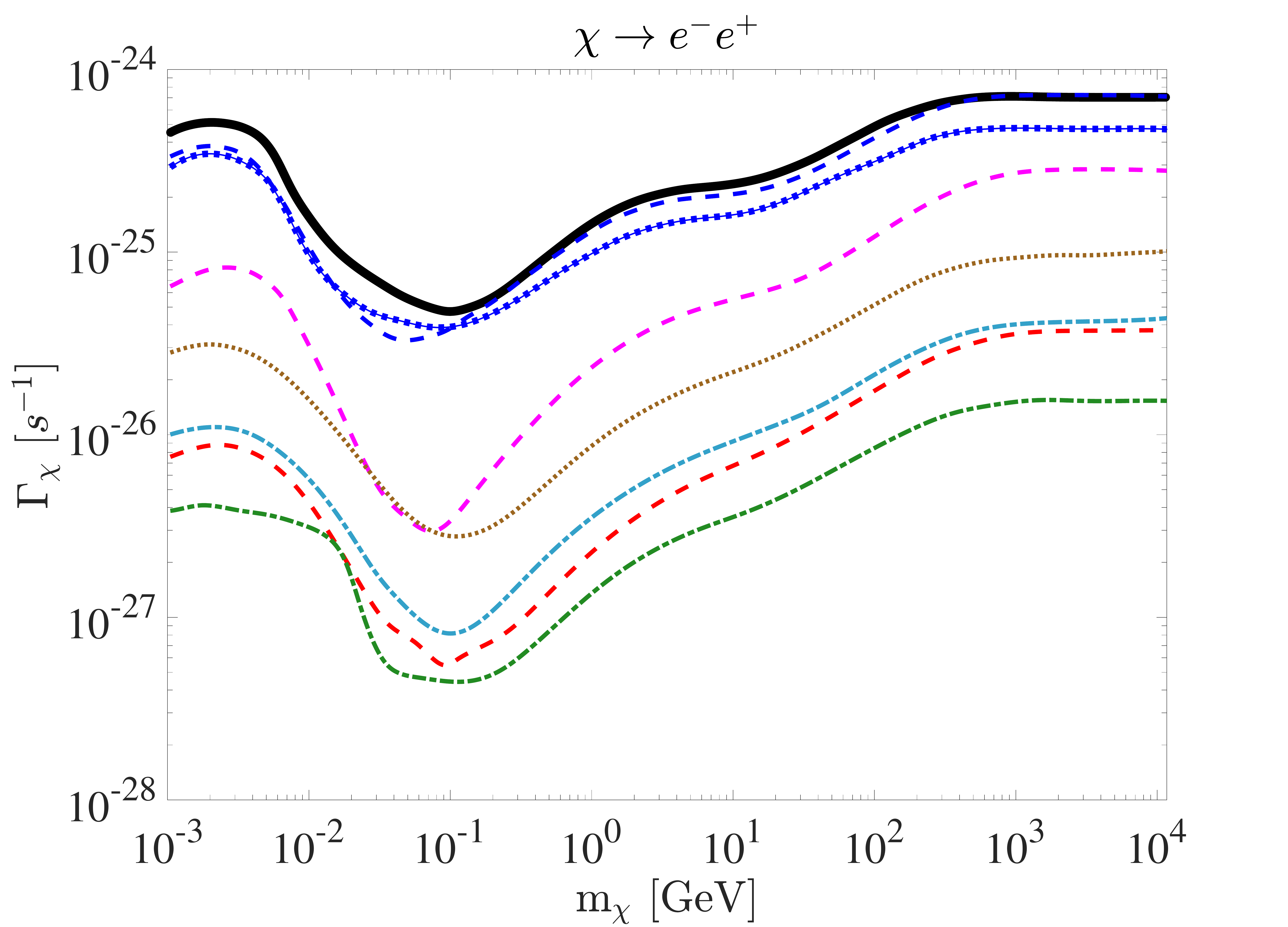}\includegraphics[width=9.5cm]{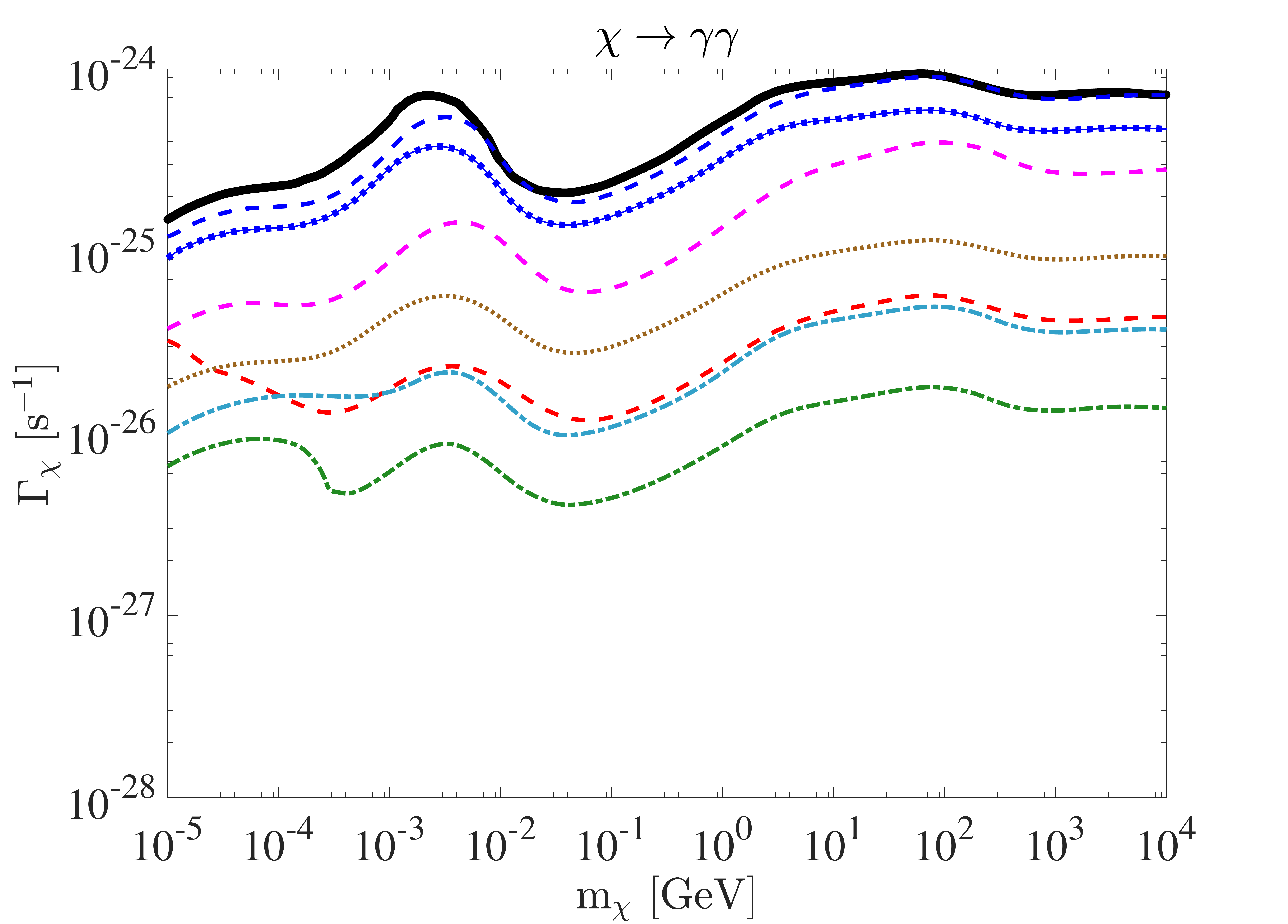}}
\subfigure{\includegraphics[width=9.5cm]{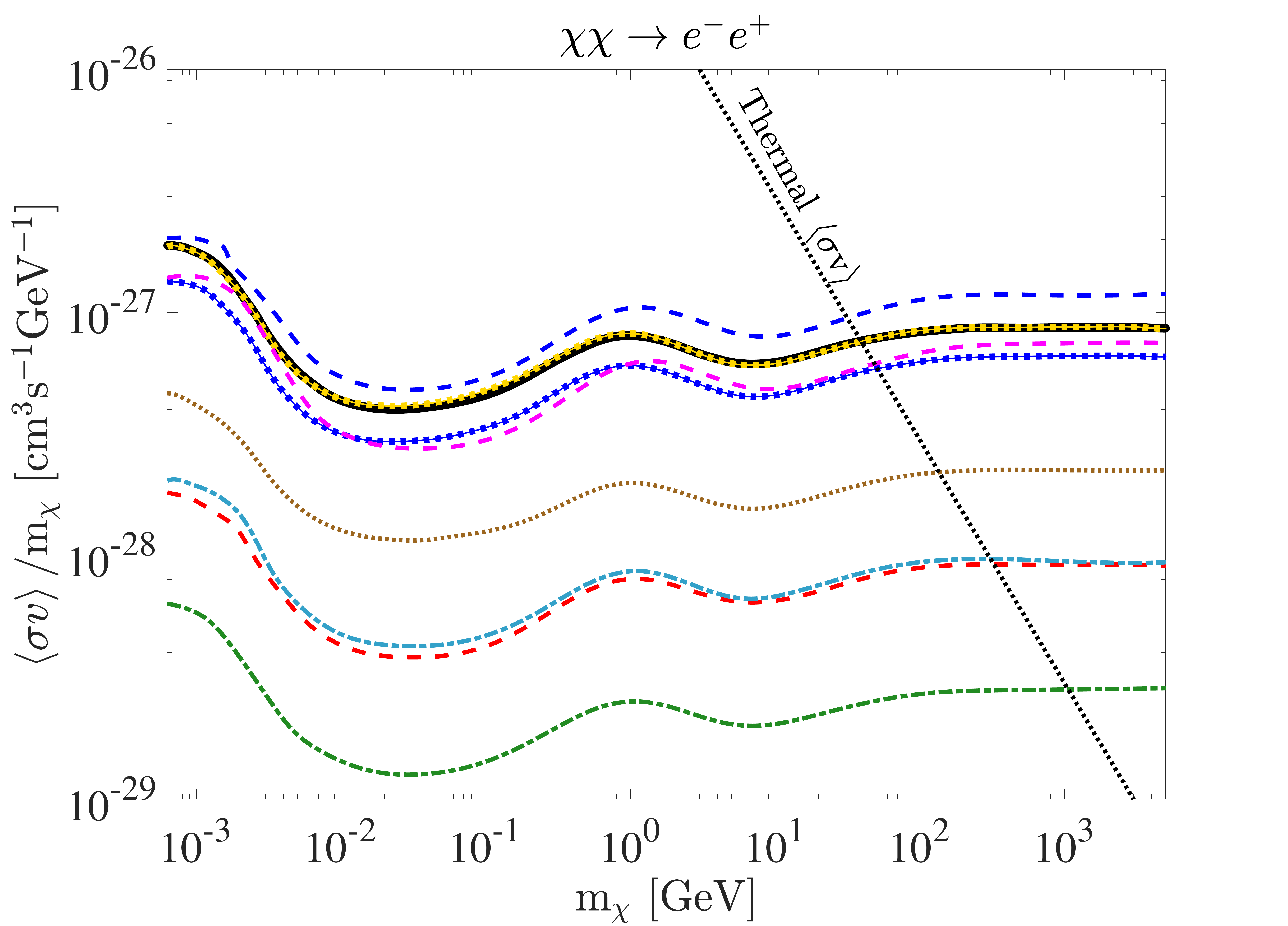}\includegraphics[width=9.5cm]{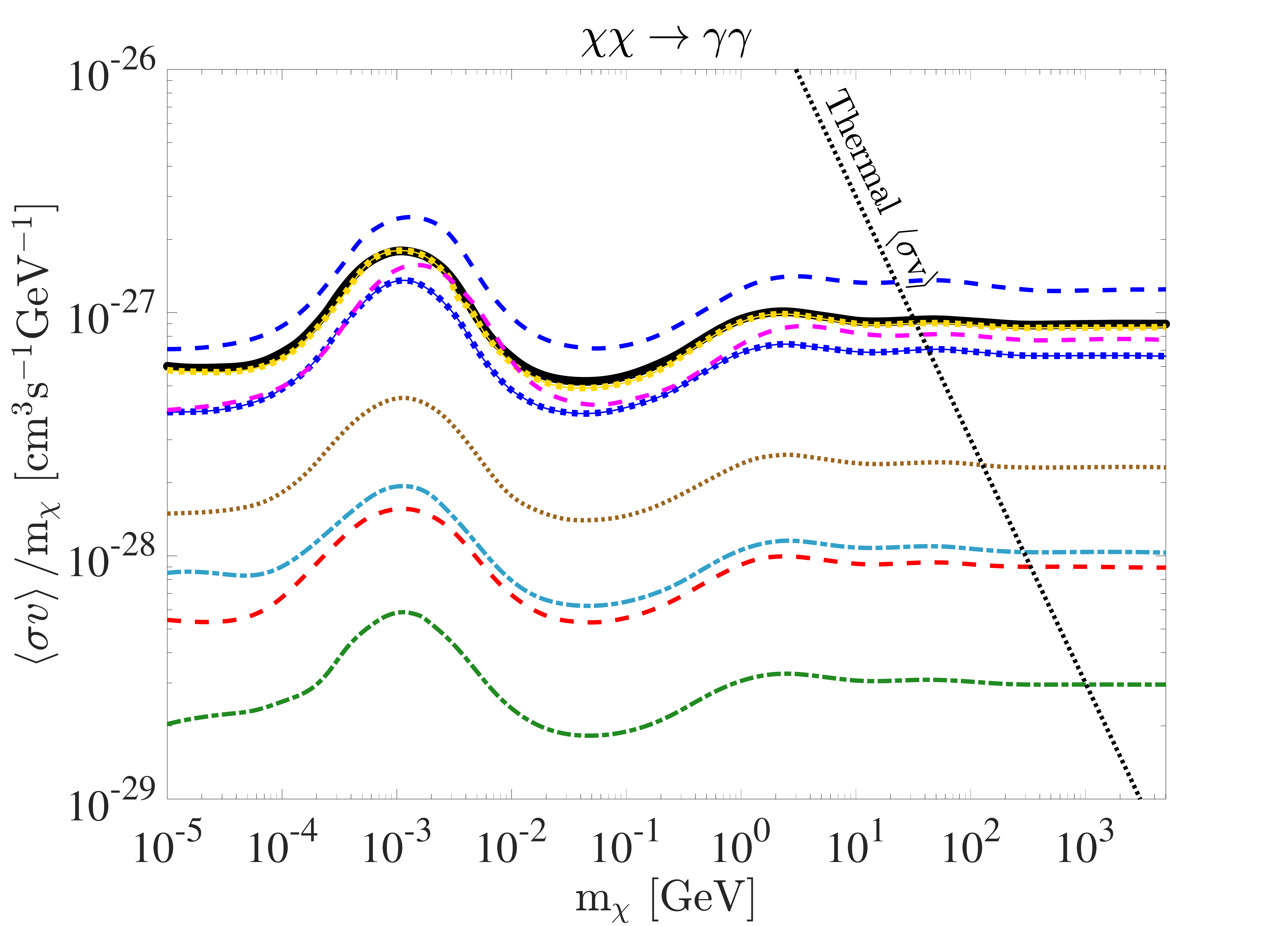}}
\subfigure{\includegraphics[width=9.5cm]{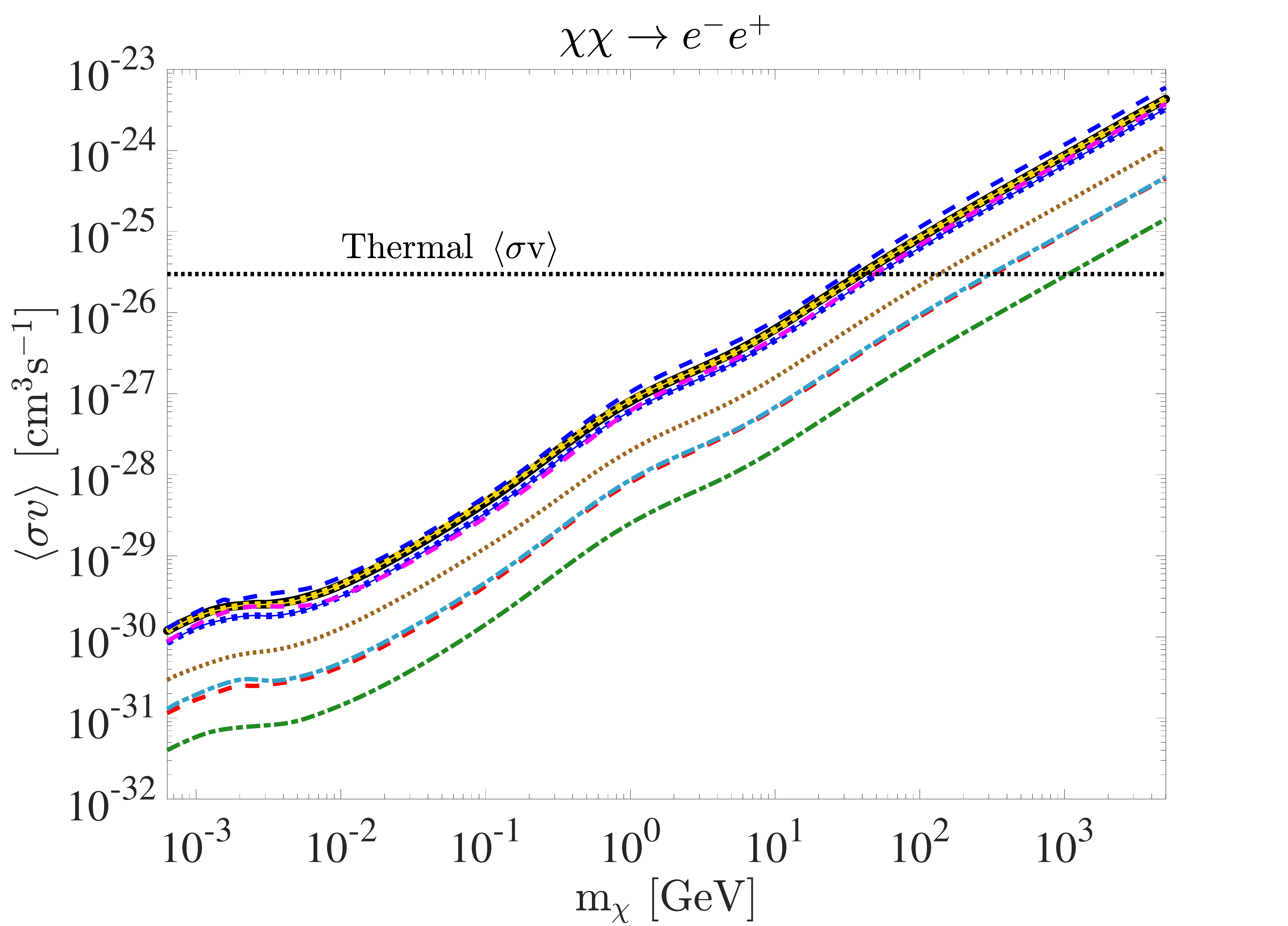}\includegraphics[width=9.5cm]{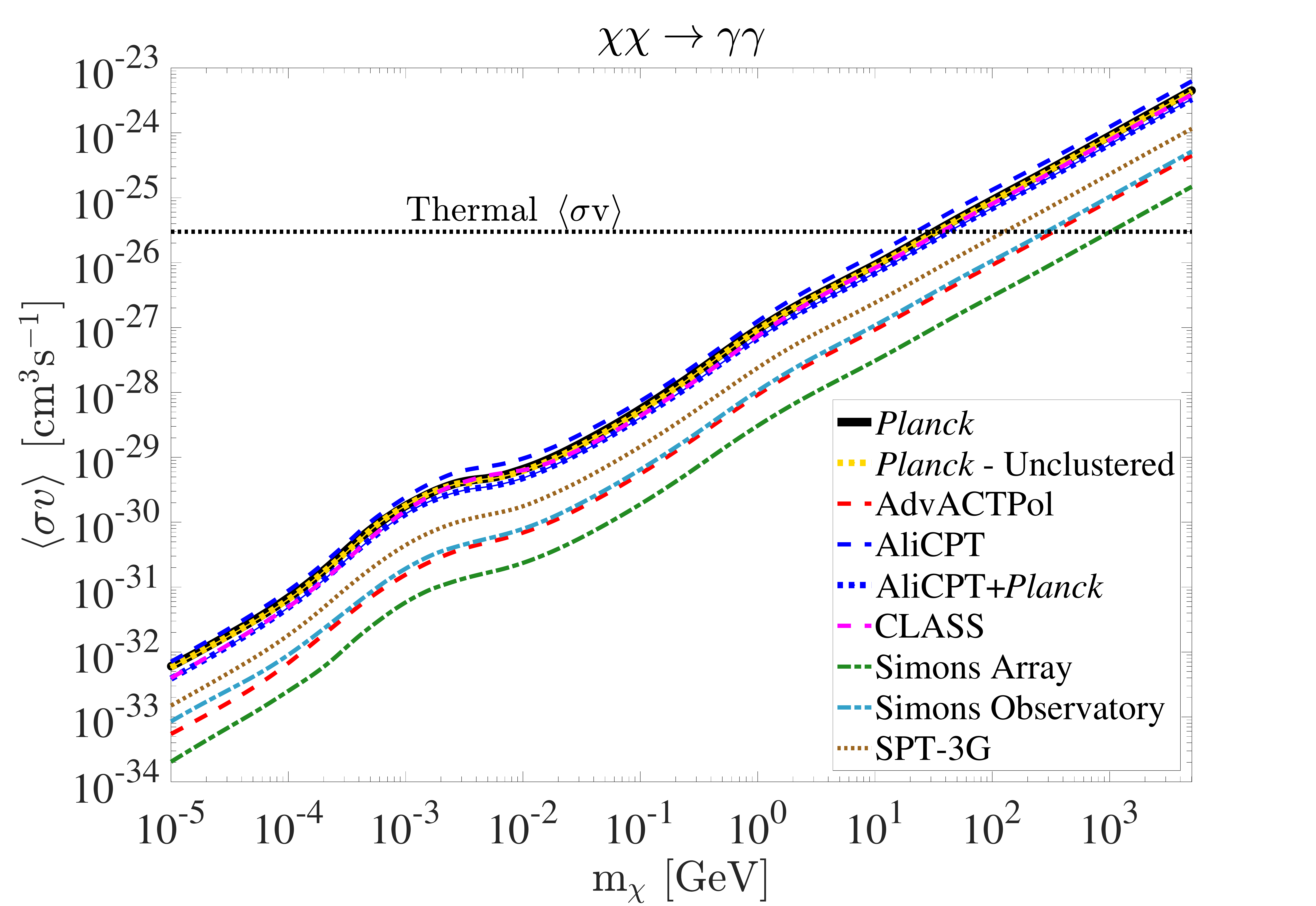}}
\caption{Limits on DM decay (top panels, $\left<\sigma v \right>/m_{\chi}=0$) and annihilation (middle and bottom panels, $\Gamma_{\chi}=0$) parameters. Regions above the lines are excluded at 95\% confidence level. All annihilation constraints assume the case with DM clustering unless labeled otherwise. Constraints labeled $Planck$ are obtained by fitting $Planck 2018$~\cite{Aghanim:2019ame} + BAO~\cite{Alam:2016hwk,Beutler:2011hx,Ross:2014qpa} datasets. Left and right panels correspond to $e^-e^+$ and $\gamma \gamma$ final states respectively. The bottom panels show the constraints in $\left<\sigma v \right>$ - ${m_{\chi}}$ plane corresponding to the $\left<\sigma v \right>/{m_{\chi}}$ -  ${m_{\chi}}$ constraints in middle panels. {\it{Planck}} constraints on annihilation in unclustered (homogeneous) DM distribution model are also shown in red dotted lines for comparison. The legend applies to all panels. $\Lambda$CDM parameters and $Planck$'s nuisance parameters are marginalized over.}
\label{fig:Result}
\end{figure*}

First we analyze the limit with {\it{Planck}} 2018~\cite{Aghanim:2019ame} and Baryon Acoustic Oscillations (BAO)~\cite{Alam:2016hwk,Beutler:2011hx,Ross:2014qpa} data as a cross-check and a benchmark. The {\it{Planck}} likelihoods used are: (i) the high-$\ell$ TTTEEE plike likelihood, (ii) low-$\ell$ TT and EE likelihoods, (iii) lensing likelihood. 
As shown in Fig.~\ref{fig:Result}, our 95\% C.L. limits for annihilation (red dotted curves) are in good agreement with Ref.~\cite{Aghanim:2018eyx} (We also found good consistency with the results in Ref.~\cite{Slatyer:2016qyl} and~\cite{Liu:2016cnk} using {\it{Planck}} 2015 data). The 21 nuisance parameters in {\it{Planck}} likelihoods are also varied in the fitting process and marginalized over in our results. 
Constraints on $\Gamma_{\chi}$ for a characteristic WIMP mass of $m_{\chi}=10{\rm{GeV}}$ are collected in Tabs.\ref{Example_Constraints_1}.
Lower bounds on $m_{\chi}$ (assuming a thermal cross section) are shown in Tab.\ref{Example_Constraints_3}.

Compared with current {\it{Planck}} limits, upcoming experiments yield either comparable or significantly improved bounds. When the limits are close to {\it{Planck}} bounds, it benefits from a joint analysis. 
For instance AliCPT sensitivity improves by around 30\% if combined with {\it{Planck}} data. 
Projected limits for AdvACTPol, SPT-3G and Simons Observatory extend to $\left<\sigma v\right>/m_{\chi}\sim\rm{10^{-28}cm^3s^{-1}GeV^{-1}}$ and $\Gamma_\chi\sim\rm{10^{-26}}s^{-1}$. 
Simons Array is estimated to be sensitive to $\left<\sigma v\right>/m_{\chi}\sim\rm{10^{-29}cm^3s^{-1}GeV^{-1}}$ and $\Gamma_\chi\sim\rm{10^{-27}}{~\rm{s^{-1}}}$. 
We also found that BICEP2/KECK Array 2018~\cite{Ade:2018gkx} places 95\% C.L. upper bounds at $\left<\sigma v\right>/m_{\chi} \sim 10^{-26} {\rm{cm^3s^{-1}GeV^{-1}}}$ and $\Gamma_{\chi} \sim 10^{-23} s^{-1}$, which are less stringent than {\it{Planck}}.
% Note : BK2 Decay Job ID: 234~237; Ann Job ID : 238, all on DM server

DM mass-dependence in the shape of sensitivity bounds reflects the energy deposit efficiency's dependence on the injection energy. The constraints are much tighter around $10$-$100$\,MeV DM mass. The reason is that electrons and photons injected at this energy range can produce photons with energies $10$\,eV-$1$\,KeV by either up-scattering on CMB photons or Compton scattering on electrons, which efficiently ionize hydrogen atoms\cite{Slatyer:2016qyl}.

The illustrated annihilation constraints assume clustering enhancements except for a comparison case (mid and lower panels, red dotted), for which we show a {\it{Planck}} bounds with a homogeneous DM distribution. The polarization anisotropy is expected to be more sensitive to higher redshift injection, around $z=600 \sim 1000$~\cite{Finkbeiner:2011dx,Contaldi:2016nys}, and the late time injection may become obscured by the uncertainty in reionization history. We found that our constraints on annihilating DM are mostly unaffected by DM clustering effect. 
For {\it {Planck}} data, inclusion of clustering enhancement gives only a slight 2\% difference in the annihilation cross-section bound. 
Given the significant $x_{\rm{e}}$ and $T_{\rm IGM}$ increase at low redshift, future precision measurement on the reionization epoch may improve this situation.

%-------------------------------------------------------------- FIG: RESULT

\begin{table}[ht]
\begin{center}
\begin{tabular}{c|c|c}
\hline
        Experiment 	      &  $\chi \to e^+ e^-$  &   $\chi \to \gamma \gamma$\\
        \hline
	{\it{Planck}}                             & $24$                    &$85$\\ 
	AdvACTPol                             & $0.68$                    &$4.7$\\
	AliCPT                                    & $21$                    &$78$\\ 
	AliCPT+{\it{Planck}}                & $16 $                    &$53$\\
	CLASS                                    & $5.5$                    &$30$\\
	Simons Array                          & $0.35$                    &$1.5 $\\
	Simons Observatory               & $0.92$                    &$4.2 $\\
	SPT-3G                                   & $2.2$                    &$9.9 $\\
\hline
\end{tabular}
\caption{
Expected 95\% C.L. upper limit on $\Gamma_{\chi}$ (in $10^{-26}{~\rm{s^{-1}}}$) at $m_{\chi}=10$ GeV.
}
\label{Example_Constraints_1}
\end{center}
\end{table}

\begin{table}[ht]
\begin{center}
\begin{tabular}{c|c|c}
\hline
	 Experiment            &$ \chi \chi \to e^+ e^-$    &$ \chi \chi \to \gamma \gamma$\\
	 \hline
	{\it{Planck}}                                   & $39$              & $32$ \\ 
	{\it{Planck}} - Unclustered            & $39$              & $33$\\
	AdvACTPol                                  & $330$              & $330$\\
	AliCPT                                          & $32$              & $22$\\
	AliCPT+{\it{Planck}}                     & $51$              & $42$ \\
	CLASS                                         & $49$              & $37$\\
	Simons Array                               & $1.1 \times 10^{3}$               & $1.0 \times 10^{3}$\\
	Simons Observatory                    & $310$              & $290$\\
	SPT-3G                                        & $140$              & $130$\\
\hline
\end{tabular}
\caption{
Expected 95\% C.L. lower limit on $m_{\chi}$ (in $\rm{GeV}$) assuming a thermal relic's annihilation cross-section $\left<\sigma v\right>=3 \times 10^{-26} {~\rm{cm^3/s}}$.
}
\label{Example_Constraints_3}
\end{center}
\end{table}

%====================================================================================   3
\section{Conclusion} \label{sec:Discussion}

WIMP dark matter decay and annihilation during the cosmic dark ages can inject high-energy particles into the intergalactic medium, which then heats and ionized the neutral gas. This effect widens the last scattering surface, attenuating polarization and temperature fluctuation on small scales while shifting peak locations of polarization anisotropy spectra.

In this paper, we made DM sensitivity forecast for several upcoming CMB experiments in detecting dark matter in 10 KeV - 10 TeV mass range that decay or annihilate into $e^-e^+$/$\gamma\gamma$. 
These experiments are either already operational or undergoing construction, including AdvACTPol, AliCPT, CLASS, Simons Array, Simons Observatory and SPT-3G. 
Assuming complete foreground removal, we found that these instruments are capable of significantly improving current CMB constraints on DM decay lifetime and annihilation cross section set by {\it{Planck}} satellite, with Simons Array giving 95\% CL upper bounds of $\left<\sigma v\right>/m_{\chi} \sim 10^{-29}{~\rm{cm}}^3{\rm{s}}^{-1}{\rm{GeV}}^{-1}$ for annihilation and $\Gamma_\chi\sim 10^{-27} {\rm{s}}^{-1}$ for decay, nearly two orders of magnitudes more stringent than {\it{Planck}} bounds. 
Assuming a thermal relic cross-section ($\left<\sigma v\right>\ =3\times 10^{-26}{~\rm{cm^3s^{-1}}}$), annihilation constraints from Simons Array can be translated into DM mass lower bounds of $m_\chi> 1.06 \rm{TeV}$ for $e^-e^+$ channel, and $m_\chi> 1.01 \rm{TeV}$ for $\gamma\gamma$ channel. 
These limits span over a wide range of DM mass that fills the energy gap in indirect cosmic X-ray and $\gamma$-ray searches and further tightens the lower bounds on the mass of thermally produced dark matter.

DM injection induces distinctly different patterns of deviation in temperature and polarization anisotropy spectra, which helps to break the degeneracy between DM and cosmological parameters. 
Promising limits on dark matter annihilation and decay rates can be expected at upcoming CMB experiments. In the case of DM annihilation, DM clustering at low redshift can have dramatic impacts on ionization fraction and IGM temperature. However, its effect on the CMB anisotropy is less prominent due to uncertainties in astrophysical injection during the reionization epoch. Future high precision CMB and 21cm experiments can significantly improve the sensitivity to DM clustering.

\medskip
{\bf Acknowledgements}

We thank Shulei Ni, Siyu Li, Hong Li, Xiaojun Bi and Steven Clark for helpful discussions, and Tracy R. Slatyer, Hongwan Liu for updated deposition efficiency data. Y.G. is supported by the Institute of High Energy Physics, Chinese Academy of Sciences under grant no.~Y7515560U1. Y.Z.M. acknowledges the supports from National Research Foundation with grant no. 105925, 109577, 120378, and 120385, and National Science Foundation China with grant no. 11828301.

\newpage
%\pagebreak[4]
%\clearpage
\begin{appendix}
\onecolumngrid
\section{Experimental Specifications}
\label{apd:Specs}

\begin{table*}[h]
\begin{center}
\begin{tabular}{lcccccc}
\toprule
\hline
${\rm{Experiment}}$  &$\ \ \ \nu{\rm{[GHz]}}\ \ \ $  &$\ \ \ \omega_{{\rm{E}},\nu}^{-1/2}\ [\mu{\rm{K}}$-${\rm{arcmin}}]\ \ \ $  &$\ \ \ \theta_{\rm{FWHM}} [\rm{arcmin}]\ \ \ $  &$\ \ \ f_{\rm{sky}}[\%]\ \ \ $  &$\ \ \ \ell_{\rm{min}}\ \ \ $   &$\ \ \ \ell_{\rm{max}}\ \ \ $\\
\hline

\midrule
\multicolumn{1}{l}{\multirow{5}{*}{AdvACTPol ~\cite{Calabrese:2014gwa,Henderson:2015nzj,Aiola:2020azj}}} 	& 28 		& 113.1    	& 7.1         &\multicolumn{1}{l}{\multirow{5}{*}{\ \ \ \ \ \ 50}}     &\multicolumn{1}{l}{\multirow{5}{*}{\ \ \ \ 350\footnote{AdvACTPol constraints would improve by a factor of 2 if choosing $\ell_{\rm{min}}=60$.}
}}     &\multicolumn{1}{l}{\multirow{5}{*}{\ \ 4000}}\\
\multicolumn{1}{l}{} 						& 41       & 99.0		& 4.8         &\multicolumn{1}{l}{}                                              &\multicolumn{1}{l}{}                                          &\multicolumn{1}{l}{}\\
\multicolumn{1}{l}{} 						& 90 $\star$       & 11.3	& 2.2         &\multicolumn{1}{l}{}                                              &\multicolumn{1}{l}{}                                          &\multicolumn{1}{l}{}\\
\multicolumn{1}{l}{} 						& 150 $\star$       & 9.9	& 1.4         &\multicolumn{1}{l}{}                                              &\multicolumn{1}{l}{}                                          &\multicolumn{1}{l}{}\\
\multicolumn{1}{l}{} 						& 230       & 35.4	& 0.9         &\multicolumn{1}{l}{}                                              &\multicolumn{1}{l}{}                                          &\multicolumn{1}{l}{}\\
\hline

\midrule
\multicolumn{1}{l}{\multirow{2}{*}{AliCPT \cite{Li:2017drr}}} 	& 90$\star$ 		& 2    	& 15.4       &\multicolumn{1}{l}{\multirow{2}{*}{\ \ \ \ \ \ 10}}     &\multicolumn{1}{l}{\multirow{2}{*}{\ \ \ \ 30}}     &\multicolumn{1}{l}{\multirow{2}{*}{\ \ \ \ 600}}\\
\multicolumn{1}{l}{} 						& 150$\star$       & 2		& 9.7         &\multicolumn{1}{l}{}                                              &\multicolumn{1}{l}{}                                          &\multicolumn{1}{l}{}\\
\hline

\midrule
\multicolumn{1}{l}{\multirow{4}{*}{CLASS  \cite{Essinger-Hileman:2014pja}}}         & 38        & 39    	& 90         &\multicolumn{1}{l}{\multirow{4}{*}{\ \ \ \ \ \ 70}}       &\multicolumn{1}{l}{\multirow{4}{*}{\ \ \ \ 5}}     &\multicolumn{1}{l}{\multirow{4}{*}{\ \ 200}}\\
\multicolumn{1}{l}{} 						 & 93$\star$         & 13 	& 40         &\multicolumn{1}{l}{}                                              &\multicolumn{1}{l}{}                                          &\multicolumn{1}{l}{}\\
\multicolumn{1}{l}{} 					 	 & 148$\star$       & 15	        & 24         &\multicolumn{1}{l}{}                                              &\multicolumn{1}{l}{}                                          &\multicolumn{1}{l}{}\\
\multicolumn{1}{l}{} 					 	 & 217       & 43	        & 18         &\multicolumn{1}{l}{}                                              &\multicolumn{1}{l}{}                                          &\multicolumn{1}{l}{}\\
\hline

\midrule
\multicolumn{1}{l}{\multirow{3}{*}{Simons Array \cite{Arnold:2014qym,Creminelli:2015oda}}}         & 95$\star$        & 13.9    	& 5.2         &\multicolumn{1}{l}{\multirow{3}{*}{\ \ \ \ \ \ 65}}       &\multicolumn{1}{l}{\multirow{3}{*}{\ \ \ \ 30}}     &\multicolumn{1}{l}{\multirow{3}{*}{\ \ 3000}}\\
\multicolumn{1}{l}{} 						 & 150$\star$       & 11.4	& 3.5         &\multicolumn{1}{l}{}                                              &\multicolumn{1}{l}{}                                          &\multicolumn{1}{l}{}\\
\multicolumn{1}{l}{} 					 	 & 220       & 30.1	        & 2.7         &\multicolumn{1}{l}{}                                              &\multicolumn{1}{l}{}                                          &\multicolumn{1}{l}{}\\
\hline

\midrule
\multicolumn{1}{l}{\multirow{6}{*}{Simons Observatory - SAT \cite{Ade:2018sbj}}} & 27 & 35.4 & 93  &\multicolumn{1}{l}{\multirow{6}{*}{\ \ \ \ \ \ 10}}     &\multicolumn{1}{l}{\multirow{6}{*}{\ \ \ \ 25}}     &\multicolumn{1}{l}{\multirow{6}{*}{\ \ 1000}}\\
\multicolumn{1}{l}{} 						 & 39         & 24	        & 63          &\multicolumn{1}{l}{}                                              &\multicolumn{1}{l}{}                                          &\multicolumn{1}{l}{}\\
\multicolumn{1}{l}{} 					 	 & 93$\star$         & 2.7	& 30         &\multicolumn{1}{l}{}                                              &\multicolumn{1}{l}{}                                          &\multicolumn{1}{l}{}\\
\multicolumn{1}{l}{} 					 	 & 145$\star$       & 3   	& 17         &\multicolumn{1}{l}{}                                              &\multicolumn{1}{l}{}                                          &\multicolumn{1}{l}{}\\
\multicolumn{1}{l}{} 					 	 & 225       & 6	        & 11          &\multicolumn{1}{l}{}                                              &\multicolumn{1}{l}{}                                          &\multicolumn{1}{l}{}\\
\multicolumn{1}{l}{} 					 	 & 280       & 14.1	& 9           &\multicolumn{1}{l}{}                                              &\multicolumn{1}{l}{}                                          &\multicolumn{1}{l}{}\\
\hline

\midrule
\multicolumn{1}{l}{\multirow{6}{*}{Simons Observatory - LAT \cite{Ade:2018sbj}}} & 27 & 73.5 & 7.4  &\multicolumn{1}{l}{\multirow{6}{*}{\ \ \ \ \ \ 40}}     &\multicolumn{1}{l}{\multirow{6}{*}{\ \ \ \ 1000}}     &\multicolumn{1}{l}{\multirow{6}{*}{\ \ 5000}}\\
\multicolumn{1}{l}{} 						 & 39         & 38.2      & 5.1          &\multicolumn{1}{l}{}                                              &\multicolumn{1}{l}{}                                          &\multicolumn{1}{l}{}\\
\multicolumn{1}{l}{} 					 	 & 93$\star$         & 8.2	& 2.2          &\multicolumn{1}{l}{}                                              &\multicolumn{1}{l}{}                                          &\multicolumn{1}{l}{}\\
\multicolumn{1}{l}{} 					 	 & 145$\star$       & 8.9   	& 1.4          &\multicolumn{1}{l}{}                                              &\multicolumn{1}{l}{}                                          &\multicolumn{1}{l}{}\\
\multicolumn{1}{l}{} 					 	 & 225       & 21.2      & 1             &\multicolumn{1}{l}{}                                              &\multicolumn{1}{l}{}                                          &\multicolumn{1}{l}{}\\
\multicolumn{1}{l}{} 					 	 & 280       & 52.3	& 0.9          &\multicolumn{1}{l}{}                                              &\multicolumn{1}{l}{}                                          &\multicolumn{1}{l}{}\\
\hline

\midrule
\multicolumn{1}{l}{\multirow{3}{*}{SPT-3G \cite{Benson:2014qhw,Anderson:2018mry,Creminelli:2015oda}}}         & 95$\star$        & 5.1    	& 1         &\multicolumn{1}{l}{\multirow{3}{*}{\ \ \ \ \ \ 6}}       &\multicolumn{1}{l}{\multirow{3}{*}{\ \ \ \ 50}}     &\multicolumn{1}{l}{\multirow{3}{*}{\ \ 5000}}\\
\multicolumn{1}{l}{} 						 & 150$\star$       & 4.7	& 1         &\multicolumn{1}{l}{}                                              &\multicolumn{1}{l}{}                                          &\multicolumn{1}{l}{}\\
\multicolumn{1}{l}{} 					 	 & 220       & 12.0	        & 1         &\multicolumn{1}{l}{}                                              &\multicolumn{1}{l}{}                                          &\multicolumn{1}{l}{}\\
\hline

%\midrule
%\multicolumn{1}{l}{\multirow{6}{*}{CMB S4 \cite{Abazajian:2019eic,Abazajian:2016yjj}}} & 30 & 30.8 & 7.4  &\multicolumn{1}{l}{\multirow{6}{*}{\ \ \ \ \ \ 62}}     &\multicolumn{1}{l}{\multirow{6}{*}{\ \ \ \ 30}}     &\multicolumn{1}{l}{\multirow{6}{*}{\ \ 4000}}\\
%\multicolumn{1}{l}{} 						 & 40         & 17.6      & 5.1          &\multicolumn{1}{l}{}                                              &\multicolumn{1}{l}{}                                          &\multicolumn{1}{l}{}\\
%\multicolumn{1}{l}{} 					 	 & 95$\star$         & 2.9	& 2.2          &\multicolumn{1}{l}{}                                              &\multicolumn{1}{l}{}                                          &\multicolumn{1}{l}{}\\
%\multicolumn{1}{l}{} 					 	 & 145$\star$       & 2.8   	& 1.4          &\multicolumn{1}{l}{}                                              &\multicolumn{1}{l}{}                                          &\multicolumn{1}{l}{}\\
%\multicolumn{1}{l}{} 					 	 & 220       & 9.8      & 1             &\multicolumn{1}{l}{}                                              &\multicolumn{1}{l}{}                                          &\multicolumn{1}{l}{}\\
%\multicolumn{1}{l}{} 					 	 & 270       & 23.6	& 0.9          &\multicolumn{1}{l}{}                                              &\multicolumn{1}{l}{}                                          &\multicolumn{1}{l}{}\\
%\hline
%
\bottomrule
\end{tabular}
\caption{Specifications for experiments considered in this work. 
Note that we only use frequency channels in $90 \sim 150 {~\rm{GHz}}$ (labeled with $\star$) range, other frequencies are usually more purposed for monitoring foregrounds and are therefore not included in our forecast.
}
%We set $\ell_{\rm{min}}$ to \ckk{50} for all ground-based experiments except for AliCPT and CLASS\cite{Essinger-Hileman:2014pja}. Maximum $\ell_{\rm{max}}$ value is set to 4000.}
\label{tab:ExpSpecs}
\end{center}
\end{table*}
\end{appendix}
%\pagebreak

\twocolumngrid
\bibliography{refs.bib}

\begin{thebibliography}{62}%
\makeatletter
\providecommand \@ifxundefined [1]{%
 \@ifx{#1\undefined}
}%
\providecommand \@ifnum [1]{%
 \ifnum #1\expandafter \@firstoftwo
 \else \expandafter \@secondoftwo
 \fi
}%
\providecommand \@ifx [1]{%
 \ifx #1\expandafter \@firstoftwo
 \else \expandafter \@secondoftwo
 \fi
}%
\providecommand \natexlab [1]{#1}%
\providecommand \enquote  [1]{``#1''}%
\providecommand \bibnamefont  [1]{#1}%
\providecommand \bibfnamefont [1]{#1}%
\providecommand \citenamefont [1]{#1}%
\providecommand \href@noop [0]{\@secondoftwo}%
\providecommand \href [0]{\begingroup \@sanitize@url \@href}%
\providecommand \@href[1]{\@@startlink{#1}\@@href}%
\providecommand \@@href[1]{\endgroup#1\@@endlink}%
\providecommand \@sanitize@url [0]{\catcode `\\12\catcode `\$12\catcode
  `\&12\catcode `\#12\catcode `\^12\catcode `\_12\catcode `\%12\relax}%
\providecommand \@@startlink[1]{}%
\providecommand \@@endlink[0]{}%
\providecommand \url  [0]{\begingroup\@sanitize@url \@url }%
\providecommand \@url [1]{\endgroup\@href {#1}{\urlprefix }}%
\providecommand \urlprefix  [0]{URL }%
\providecommand \Eprint [0]{\href }%
\providecommand \doibase [0]{http://dx.doi.org/}%
\providecommand \selectlanguage [0]{\@gobble}%
\providecommand \bibinfo  [0]{\@secondoftwo}%
\providecommand \bibfield  [0]{\@secondoftwo}%
\providecommand \translation [1]{[#1]}%
\providecommand \BibitemOpen [0]{}%
\providecommand \bibitemStop [0]{}%
\providecommand \bibitemNoStop [0]{.\EOS\space}%
\providecommand \EOS [0]{\spacefactor3000\relax}%
\providecommand \BibitemShut  [1]{\csname bibitem#1\endcsname}%
\let\auto@bib@innerbib\@empty
%</preamble>
\bibitem [{\citenamefont {Cushman}\ \emph {et~al.}(2013)\citenamefont {Cushman}
  \emph {et~al.}}]{Cushman:2013zza}%
  \BibitemOpen
  \bibfield  {author} {\bibinfo {author} {\bibfnamefont {P.}~\bibnamefont
  {Cushman}} \emph {et~al.},\ }in\ \href@noop {} {\emph {\bibinfo {booktitle}
  {Community Summer Study 2013}}}\ (\bibinfo {year} {2013})\ \Eprint
  {http://arxiv.org/abs/1310.8327} {arXiv:1310.8327 [hep-ex]} \BibitemShut
  {NoStop}%
\bibitem [{\citenamefont {Profumo}(2013)}]{Profumo:2013yn}%
  \BibitemOpen
  \bibfield  {author} {\bibinfo {author} {\bibfnamefont {S.}~\bibnamefont
  {Profumo}}\ }(\bibinfo {year} {2013})\ pp.\ \bibinfo {pages} {143--189},\
  \Eprint {http://arxiv.org/abs/1301.0952} {arXiv:1301.0952 [hep-ph]}
  \BibitemShut {NoStop}%
\bibitem [{\citenamefont {Boveia}\ and\ \citenamefont
  {Doglioni}(2018)}]{Boveia:2018yeb}%
  \BibitemOpen
  \bibfield  {author} {\bibinfo {author} {\bibfnamefont {A.}~\bibnamefont
  {Boveia}}\ and\ \bibinfo {author} {\bibfnamefont {C.}~\bibnamefont
  {Doglioni}},\ }\href {\doibase 10.1146/annurev-nucl-101917-021008} {\bibfield
   {journal} {\bibinfo  {journal} {Ann. Rev. Nucl. Part. Sci.}\ }\textbf
  {\bibinfo {volume} {68}},\ \bibinfo {pages} {429} (\bibinfo {year} {2018})},\
  \Eprint {http://arxiv.org/abs/1810.12238} {arXiv:1810.12238 [hep-ex]}
  \BibitemShut {NoStop}%
%%CITATION = ARXIV:1810.12238;%%
\bibitem [{\citenamefont {Schumann}(2019)}]{Schumann:2019eaa}%
  \BibitemOpen
  \bibfield  {author} {\bibinfo {author} {\bibfnamefont {M.}~\bibnamefont
  {Schumann}},\ }\href {\doibase 10.1088/1361-6471/ab2ea5} {\bibfield
  {journal} {\bibinfo  {journal} {J. Phys.}\ }\textbf {\bibinfo {volume}
  {G46}},\ \bibinfo {pages} {103003} (\bibinfo {year} {2019})},\ \Eprint
  {http://arxiv.org/abs/1903.03026} {arXiv:1903.03026 [astro-ph.CO]}
  \BibitemShut {NoStop}%
%%CITATION = ARXIV:1903.03026;%%
\bibitem [{\citenamefont {Chen}\ and\ \citenamefont
  {Kamionkowski}(2004)}]{Chen:2003gz}%
  \BibitemOpen
  \bibfield  {author} {\bibinfo {author} {\bibfnamefont {X.-L.}\ \bibnamefont
  {Chen}}\ and\ \bibinfo {author} {\bibfnamefont {M.}~\bibnamefont
  {Kamionkowski}},\ }\href {\doibase 10.1103/PhysRevD.70.043502} {\bibfield
  {journal} {\bibinfo  {journal} {Phys. Rev.}\ }\textbf {\bibinfo {volume}
  {D70}},\ \bibinfo {pages} {043502} (\bibinfo {year} {2004})},\ \Eprint
  {http://arxiv.org/abs/astro-ph/0310473} {arXiv:astro-ph/0310473 [astro-ph]}
  \BibitemShut {NoStop}%
%%CITATION = ASTRO-PH/0310473;%%
\bibitem [{\citenamefont {Adams}\ \emph {et~al.}(1998)\citenamefont {Adams},
  \citenamefont {Sarkar},\ and\ \citenamefont {Sciama}}]{Adams:1998nr}%
  \BibitemOpen
  \bibfield  {author} {\bibinfo {author} {\bibfnamefont {J.~A.}\ \bibnamefont
  {Adams}}, \bibinfo {author} {\bibfnamefont {S.}~\bibnamefont {Sarkar}}, \
  and\ \bibinfo {author} {\bibfnamefont {D.}~\bibnamefont {Sciama}},\ }\href
  {\doibase 10.1046/j.1365-8711.1998.02017.x} {\bibfield  {journal} {\bibinfo
  {journal} {Mon. Not. Roy. Astron. Soc.}\ }\textbf {\bibinfo {volume} {301}},\
  \bibinfo {pages} {210} (\bibinfo {year} {1998})},\ \Eprint
  {http://arxiv.org/abs/astro-ph/9805108} {arXiv:astro-ph/9805108} \BibitemShut
  {NoStop}%
\bibitem [{\citenamefont {Padmanabhan}\ and\ \citenamefont
  {Finkbeiner}(2005)}]{Padmanabhan:2005es}%
  \BibitemOpen
  \bibfield  {author} {\bibinfo {author} {\bibfnamefont {N.}~\bibnamefont
  {Padmanabhan}}\ and\ \bibinfo {author} {\bibfnamefont {D.~P.}\ \bibnamefont
  {Finkbeiner}},\ }\href {\doibase 10.1103/PhysRevD.72.023508} {\bibfield
  {journal} {\bibinfo  {journal} {Phys. Rev.}\ }\textbf {\bibinfo {volume}
  {D72}},\ \bibinfo {pages} {023508} (\bibinfo {year} {2005})},\ \Eprint
  {http://arxiv.org/abs/astro-ph/0503486} {arXiv:astro-ph/0503486 [astro-ph]}
  \BibitemShut {NoStop}%
%%CITATION = ASTRO-PH/0503486;%%
\bibitem [{\citenamefont {Pierpaoli}(2004)}]{Pierpaoli:2003rz}%
  \BibitemOpen
  \bibfield  {author} {\bibinfo {author} {\bibfnamefont {E.}~\bibnamefont
  {Pierpaoli}},\ }\href {\doibase 10.1103/PhysRevLett.92.031301} {\bibfield
  {journal} {\bibinfo  {journal} {Phys. Rev. Lett.}\ }\textbf {\bibinfo
  {volume} {92}},\ \bibinfo {pages} {031301} (\bibinfo {year} {2004})},\
  \Eprint {http://arxiv.org/abs/astro-ph/0310375} {arXiv:astro-ph/0310375
  [astro-ph]} \BibitemShut {NoStop}%
%%CITATION = ASTRO-PH/0310375;%%
\bibitem [{\citenamefont {Acharya}\ and\ \citenamefont
  {Khatri}(2019)}]{Acharya:2019uba}%
  \BibitemOpen
  \bibfield  {author} {\bibinfo {author} {\bibfnamefont {S.~K.}\ \bibnamefont
  {Acharya}}\ and\ \bibinfo {author} {\bibfnamefont {R.}~\bibnamefont
  {Khatri}},\ }\href@noop {} {\  (\bibinfo {year} {2019})},\ \Eprint
  {http://arxiv.org/abs/1910.06272} {arXiv:1910.06272 [astro-ph.CO]}
  \BibitemShut {NoStop}%
%%CITATION = ARXIV:1910.06272;%%
\bibitem [{\citenamefont {Aghanim}\ \emph {et~al.}(2018)\citenamefont {Aghanim}
  \emph {et~al.}}]{Aghanim:2018eyx}%
  \BibitemOpen
  \bibfield  {author} {\bibinfo {author} {\bibfnamefont {N.}~\bibnamefont
  {Aghanim}} \emph {et~al.} (\bibinfo {collaboration} {Planck}),\ }\href@noop
  {} {\  (\bibinfo {year} {2018})},\ \Eprint {http://arxiv.org/abs/1807.06209}
  {arXiv:1807.06209 [astro-ph.CO]} \BibitemShut {NoStop}%
%%CITATION = ARXIV:1807.06209;%%
\bibitem [{\citenamefont {Slatyer}\ and\ \citenamefont
  {Wu}(2017)}]{Slatyer:2016qyl}%
  \BibitemOpen
  \bibfield  {author} {\bibinfo {author} {\bibfnamefont {T.~R.}\ \bibnamefont
  {Slatyer}}\ and\ \bibinfo {author} {\bibfnamefont {C.-L.}\ \bibnamefont
  {Wu}},\ }\href {\doibase 10.1103/PhysRevD.95.023010} {\bibfield  {journal}
  {\bibinfo  {journal} {Phys. Rev.}\ }\textbf {\bibinfo {volume} {D95}},\
  \bibinfo {pages} {023010} (\bibinfo {year} {2017})},\ \Eprint
  {http://arxiv.org/abs/1610.06933} {arXiv:1610.06933 [astro-ph.CO]}
  \BibitemShut {NoStop}%
%%CITATION = ARXIV:1610.06933;%%
\bibitem [{\citenamefont {Slatyer}(2016{\natexlab{a}})}]{Slatyer:2015jla}%
  \BibitemOpen
  \bibfield  {author} {\bibinfo {author} {\bibfnamefont {T.~R.}\ \bibnamefont
  {Slatyer}},\ }\href {\doibase 10.1103/PhysRevD.93.023527} {\bibfield
  {journal} {\bibinfo  {journal} {Phys. Rev.}\ }\textbf {\bibinfo {volume}
  {D93}},\ \bibinfo {pages} {023527} (\bibinfo {year} {2016}{\natexlab{a}})},\
  \Eprint {http://arxiv.org/abs/1506.03811} {arXiv:1506.03811 [hep-ph]}
  \BibitemShut {NoStop}%
%%CITATION = ARXIV:1506.03811;%%
\bibitem [{\citenamefont {Liu}\ \emph {et~al.}(2016)\citenamefont {Liu},
  \citenamefont {Slatyer},\ and\ \citenamefont {Zavala}}]{Liu:2016cnk}%
  \BibitemOpen
  \bibfield  {author} {\bibinfo {author} {\bibfnamefont {H.}~\bibnamefont
  {Liu}}, \bibinfo {author} {\bibfnamefont {T.~R.}\ \bibnamefont {Slatyer}}, \
  and\ \bibinfo {author} {\bibfnamefont {J.}~\bibnamefont {Zavala}},\ }\href
  {\doibase 10.1103/PhysRevD.94.063507} {\bibfield  {journal} {\bibinfo
  {journal} {Phys. Rev.}\ }\textbf {\bibinfo {volume} {D94}},\ \bibinfo {pages}
  {063507} (\bibinfo {year} {2016})},\ \Eprint
  {http://arxiv.org/abs/1604.02457} {arXiv:1604.02457 [astro-ph.CO]}
  \BibitemShut {NoStop}%
%%CITATION = ARXIV:1604.02457;%%
\bibitem [{\citenamefont {Kawasaki}\ \emph {et~al.}(2016)\citenamefont
  {Kawasaki}, \citenamefont {Nakayama},\ and\ \citenamefont
  {Sekiguchi}}]{Kawasaki:2015peu}%
  \BibitemOpen
  \bibfield  {author} {\bibinfo {author} {\bibfnamefont {M.}~\bibnamefont
  {Kawasaki}}, \bibinfo {author} {\bibfnamefont {K.}~\bibnamefont {Nakayama}},
  \ and\ \bibinfo {author} {\bibfnamefont {T.}~\bibnamefont {Sekiguchi}},\
  }\href {\doibase 10.1016/j.physletb.2016.03.005} {\bibfield  {journal}
  {\bibinfo  {journal} {Phys. Lett.}\ }\textbf {\bibinfo {volume} {B756}},\
  \bibinfo {pages} {212} (\bibinfo {year} {2016})},\ \Eprint
  {http://arxiv.org/abs/1512.08015} {arXiv:1512.08015 [astro-ph.CO]}
  \BibitemShut {NoStop}%
%%CITATION = ARXIV:1512.08015;%%
\bibitem [{\citenamefont {Aghanim}\ \emph {et~al.}(2019)\citenamefont {Aghanim}
  \emph {et~al.}}]{Aghanim:2019ame}%
  \BibitemOpen
  \bibfield  {author} {\bibinfo {author} {\bibfnamefont {N.}~\bibnamefont
  {Aghanim}} \emph {et~al.} (\bibinfo {collaboration} {Planck}),\ }\href@noop
  {} {\  (\bibinfo {year} {2019})},\ \Eprint {http://arxiv.org/abs/1907.12875}
  {arXiv:1907.12875 [astro-ph.CO]} \BibitemShut {NoStop}%
%%CITATION = ARXIV:1907.12875;%%
\bibitem [{\citenamefont {Aghanim}\ \emph {et~al.}(2016)\citenamefont {Aghanim}
  \emph {et~al.}}]{Aghanim:2015xee}%
  \BibitemOpen
  \bibfield  {author} {\bibinfo {author} {\bibfnamefont {N.}~\bibnamefont
  {Aghanim}} \emph {et~al.} (\bibinfo {collaboration} {Planck}),\ }\href
  {\doibase 10.1051/0004-6361/201526926} {\bibfield  {journal} {\bibinfo
  {journal} {Astron. Astrophys.}\ }\textbf {\bibinfo {volume} {594}},\ \bibinfo
  {pages} {A11} (\bibinfo {year} {2016})},\ \Eprint
  {http://arxiv.org/abs/1507.02704} {arXiv:1507.02704 [astro-ph.CO]}
  \BibitemShut {NoStop}%
%%CITATION = ARXIV:1507.02704;%%
\bibitem [{\citenamefont {Ahnen}\ \emph {et~al.}(2016)\citenamefont {Ahnen}
  \emph {et~al.}}]{Ahnen:2016qkx}%
  \BibitemOpen
  \bibfield  {author} {\bibinfo {author} {\bibfnamefont {M.~L.}\ \bibnamefont
  {Ahnen}} \emph {et~al.} (\bibinfo {collaboration} {MAGIC, Fermi-LAT}),\
  }\href {\doibase 10.1088/1475-7516/2016/02/039} {\bibfield  {journal}
  {\bibinfo  {journal} {JCAP}\ }\textbf {\bibinfo {volume} {1602}},\ \bibinfo
  {pages} {039} (\bibinfo {year} {2016})},\ \Eprint
  {http://arxiv.org/abs/1601.06590} {arXiv:1601.06590 [astro-ph.HE]}
  \BibitemShut {NoStop}%
%%CITATION = ARXIV:1601.06590;%%
\bibitem [{\citenamefont {Grayson}\ \emph {et~al.}(2016)\citenamefont {Grayson}
  \emph {et~al.}}]{Grayson:2016smb}%
  \BibitemOpen
  \bibfield  {author} {\bibinfo {author} {\bibfnamefont {J.~A.}\ \bibnamefont
  {Grayson}} \emph {et~al.} (\bibinfo {collaboration} {BICEP3}),\ }\href
  {\doibase 10.1117/12.2233894} {\bibfield  {journal} {\bibinfo  {journal}
  {Proc. SPIE Int. Soc. Opt. Eng.}\ }\textbf {\bibinfo {volume} {9914}},\
  \bibinfo {pages} {99140S} (\bibinfo {year} {2016})},\ \Eprint
  {http://arxiv.org/abs/1607.04668} {arXiv:1607.04668 [astro-ph.IM]}
  \BibitemShut {NoStop}%
%%CITATION = ARXIV:1607.04668;%%
\bibitem [{\citenamefont {Anderson}\ \emph {et~al.}(2018)\citenamefont
  {Anderson} \emph {et~al.}}]{Anderson:2018mry}%
  \BibitemOpen
  \bibfield  {author} {\bibinfo {author} {\bibfnamefont {A.~J.}\ \bibnamefont
  {Anderson}} \emph {et~al.} (\bibinfo {collaboration} {SPT}),\ }\href
  {\doibase 10.1007/s10909-018-2007-z} {\bibfield  {journal} {\bibinfo
  {journal} {J. Low. Temp. Phys.}\ }\textbf {\bibinfo {volume} {193}},\
  \bibinfo {pages} {1057} (\bibinfo {year} {2018})}\BibitemShut {NoStop}%
%%CITATION = JLTPA,193,1057;%%
\bibitem [{\citenamefont {Calabrese}\ \emph {et~al.}(2014)\citenamefont
  {Calabrese} \emph {et~al.}}]{Calabrese:2014gwa}%
  \BibitemOpen
  \bibfield  {author} {\bibinfo {author} {\bibfnamefont {E.}~\bibnamefont
  {Calabrese}} \emph {et~al.},\ }\href {\doibase 10.1088/1475-7516/2014/08/010}
  {\bibfield  {journal} {\bibinfo  {journal} {JCAP}\ }\textbf {\bibinfo
  {volume} {1408}},\ \bibinfo {pages} {010} (\bibinfo {year} {2014})},\ \Eprint
  {http://arxiv.org/abs/1406.4794} {arXiv:1406.4794 [astro-ph.CO]} \BibitemShut
  {NoStop}%
%%CITATION = ARXIV:1406.4794;%%
\bibitem [{\citenamefont {Li}\ \emph {et~al.}(2018{\natexlab{a}})\citenamefont
  {Li} \emph {et~al.}}]{Li:2018uwb}%
  \BibitemOpen
  \bibfield  {author} {\bibinfo {author} {\bibfnamefont {Y.}~\bibnamefont {Li}}
  \emph {et~al.},\ }\href {\doibase 10.1117/12.2313942} {\bibfield  {journal}
  {\bibinfo  {journal} {Proc. SPIE Int. Soc. Opt. Eng.}\ }\textbf {\bibinfo
  {volume} {10708}},\ \bibinfo {pages} {107080A} (\bibinfo {year}
  {2018}{\natexlab{a}})}\BibitemShut {NoStop}%
%%CITATION = PSISD,10708,107080A;%%
\bibitem [{\citenamefont {Essinger-Hileman}\ \emph {et~al.}(2014)\citenamefont
  {Essinger-Hileman} \emph {et~al.}}]{Essinger-Hileman:2014pja}%
  \BibitemOpen
  \bibfield  {author} {\bibinfo {author} {\bibfnamefont {T.}~\bibnamefont
  {Essinger-Hileman}} \emph {et~al.},\ }\href {\doibase 10.1117/12.2056701}
  {\bibfield  {journal} {\bibinfo  {journal} {Proc. SPIE Int. Soc. Opt. Eng.}\
  }\textbf {\bibinfo {volume} {9153}},\ \bibinfo {pages} {91531I} (\bibinfo
  {year} {2014})},\ \Eprint {http://arxiv.org/abs/1408.4788} {arXiv:1408.4788
  [astro-ph.IM]} \BibitemShut {NoStop}%
%%CITATION = ARXIV:1408.4788;%%
\bibitem [{\citenamefont {Suzuki}\ \emph {et~al.}(2016)\citenamefont {Suzuki}
  \emph {et~al.}}]{Suzuki:2015zzg}%
  \BibitemOpen
  \bibfield  {author} {\bibinfo {author} {\bibfnamefont {A.}~\bibnamefont
  {Suzuki}} \emph {et~al.} (\bibinfo {collaboration} {POLARBEAR}),\ }\href
  {\doibase 10.1007/s10909-015-1425-4} {\bibfield  {journal} {\bibinfo
  {journal} {J. Low. Temp. Phys.}\ }\textbf {\bibinfo {volume} {184}},\
  \bibinfo {pages} {805} (\bibinfo {year} {2016})},\ \Eprint
  {http://arxiv.org/abs/1512.07299} {arXiv:1512.07299 [astro-ph.IM]}
  \BibitemShut {NoStop}%
%%CITATION = ARXIV:1512.07299;%%
\bibitem [{\citenamefont {Arnold}\ \emph {et~al.}(2014)\citenamefont {Arnold}
  \emph {et~al.}}]{Arnold:2014qym}%
  \BibitemOpen
  \bibfield  {author} {\bibinfo {author} {\bibfnamefont {K.}~\bibnamefont
  {Arnold}} \emph {et~al.},\ }\href {\doibase 10.1117/12.2057332} {\bibfield
  {journal} {\bibinfo  {journal} {Proc. SPIE Int. Soc. Opt. Eng.}\ }\textbf
  {\bibinfo {volume} {9153}},\ \bibinfo {pages} {91531F} (\bibinfo {year}
  {2014})}\BibitemShut {NoStop}%
%%CITATION = PSISD,9153,91531F;%%
\bibitem [{\citenamefont {Ade}\ \emph {et~al.}(2019)\citenamefont {Ade} \emph
  {et~al.}}]{Ade:2018sbj}%
  \BibitemOpen
  \bibfield  {author} {\bibinfo {author} {\bibfnamefont {P.}~\bibnamefont
  {Ade}} \emph {et~al.} (\bibinfo {collaboration} {Simons Observatory}),\
  }\href {\doibase 10.1088/1475-7516/2019/02/056} {\bibfield  {journal}
  {\bibinfo  {journal} {JCAP}\ }\textbf {\bibinfo {volume} {1902}},\ \bibinfo
  {pages} {056} (\bibinfo {year} {2019})},\ \Eprint
  {http://arxiv.org/abs/1808.07445} {arXiv:1808.07445 [astro-ph.CO]}
  \BibitemShut {NoStop}%
%%CITATION = ARXIV:1808.07445;%%
\bibitem [{\citenamefont {Li}\ \emph {et~al.}(2017)\citenamefont {Li},
  \citenamefont {Liu}, \citenamefont {Li}, \citenamefont {Li},\ and\
  \citenamefont {Zhang}}]{Li:2017lat}%
  \BibitemOpen
  \bibfield  {author} {\bibinfo {author} {\bibfnamefont {Y.-P.}\ \bibnamefont
  {Li}}, \bibinfo {author} {\bibfnamefont {Y.}~\bibnamefont {Liu}}, \bibinfo
  {author} {\bibfnamefont {S.-Y.}\ \bibnamefont {Li}}, \bibinfo {author}
  {\bibfnamefont {H.}~\bibnamefont {Li}}, \ and\ \bibinfo {author}
  {\bibfnamefont {X.}~\bibnamefont {Zhang}},\ }\href@noop {} {\  (\bibinfo
  {year} {2017})},\ \Eprint {http://arxiv.org/abs/1709.09053} {arXiv:1709.09053
  [astro-ph.IM]} \BibitemShut {NoStop}%
%%CITATION = ARXIV:1709.09053;%%
\bibitem [{\citenamefont {Weniger}\ \emph {et~al.}(2013)\citenamefont
  {Weniger}, \citenamefont {Serpico}, \citenamefont {Iocco},\ and\
  \citenamefont {Bertone}}]{Weniger:2013hja}%
  \BibitemOpen
  \bibfield  {author} {\bibinfo {author} {\bibfnamefont {C.}~\bibnamefont
  {Weniger}}, \bibinfo {author} {\bibfnamefont {P.~D.}\ \bibnamefont
  {Serpico}}, \bibinfo {author} {\bibfnamefont {F.}~\bibnamefont {Iocco}}, \
  and\ \bibinfo {author} {\bibfnamefont {G.}~\bibnamefont {Bertone}},\ }\href
  {\doibase 10.1103/PhysRevD.87.123008} {\bibfield  {journal} {\bibinfo
  {journal} {Phys. Rev.}\ }\textbf {\bibinfo {volume} {D87}},\ \bibinfo {pages}
  {123008} (\bibinfo {year} {2013})},\ \Eprint {http://arxiv.org/abs/1303.0942}
  {arXiv:1303.0942 [astro-ph.CO]} \BibitemShut {NoStop}%
%%CITATION = ARXIV:1303.0942;%%
\bibitem [{\citenamefont {Slatyer}\ \emph {et~al.}(2009)\citenamefont
  {Slatyer}, \citenamefont {Padmanabhan},\ and\ \citenamefont
  {Finkbeiner}}]{Slatyer:2009yq}%
  \BibitemOpen
  \bibfield  {author} {\bibinfo {author} {\bibfnamefont {T.~R.}\ \bibnamefont
  {Slatyer}}, \bibinfo {author} {\bibfnamefont {N.}~\bibnamefont
  {Padmanabhan}}, \ and\ \bibinfo {author} {\bibfnamefont {D.~P.}\ \bibnamefont
  {Finkbeiner}},\ }\href {\doibase 10.1103/PhysRevD.80.043526} {\bibfield
  {journal} {\bibinfo  {journal} {Phys. Rev.}\ }\textbf {\bibinfo {volume}
  {D80}},\ \bibinfo {pages} {043526} (\bibinfo {year} {2009})},\ \Eprint
  {http://arxiv.org/abs/0906.1197} {arXiv:0906.1197 [astro-ph.CO]} \BibitemShut
  {NoStop}%
%%CITATION = ARXIV:0906.1197;%%
\bibitem [{\citenamefont {Slatyer}(2013)}]{Slatyer:2012yq}%
  \BibitemOpen
  \bibfield  {author} {\bibinfo {author} {\bibfnamefont {T.~R.}\ \bibnamefont
  {Slatyer}},\ }\href {\doibase 10.1103/PhysRevD.87.123513} {\bibfield
  {journal} {\bibinfo  {journal} {Phys. Rev.}\ }\textbf {\bibinfo {volume}
  {D87}},\ \bibinfo {pages} {123513} (\bibinfo {year} {2013})},\ \Eprint
  {http://arxiv.org/abs/1211.0283} {arXiv:1211.0283 [astro-ph.CO]} \BibitemShut
  {NoStop}%
%%CITATION = ARXIV:1211.0283;%%
\bibitem [{\citenamefont {Belotsky}\ and\ \citenamefont
  {Kirillov}(2015)}]{Belotsky:2014twa}%
  \BibitemOpen
  \bibfield  {author} {\bibinfo {author} {\bibfnamefont {K.~M.}\ \bibnamefont
  {Belotsky}}\ and\ \bibinfo {author} {\bibfnamefont {A.~A.}\ \bibnamefont
  {Kirillov}},\ }\href {\doibase 10.1088/1475-7516/2015/01/041} {\bibfield
  {journal} {\bibinfo  {journal} {JCAP}\ }\textbf {\bibinfo {volume} {1501}},\
  \bibinfo {pages} {041} (\bibinfo {year} {2015})},\ \Eprint
  {http://arxiv.org/abs/1409.8601} {arXiv:1409.8601 [astro-ph.CO]} \BibitemShut
  {NoStop}%
%%CITATION = ARXIV:1409.8601;%%
\bibitem [{\citenamefont {Liu}\ \emph {et~al.}(2019)\citenamefont {Liu},
  \citenamefont {Ridgway},\ and\ \citenamefont {Slatyer}}]{Liu:2019bbm}%
  \BibitemOpen
  \bibfield  {author} {\bibinfo {author} {\bibfnamefont {H.}~\bibnamefont
  {Liu}}, \bibinfo {author} {\bibfnamefont {G.~W.}\ \bibnamefont {Ridgway}}, \
  and\ \bibinfo {author} {\bibfnamefont {T.~R.}\ \bibnamefont {Slatyer}},\
  }\href@noop {} {\  (\bibinfo {year} {2019})},\ \Eprint
  {http://arxiv.org/abs/1904.09296} {arXiv:1904.09296 [astro-ph.CO]}
  \BibitemShut {NoStop}%
%%CITATION = ARXIV:1904.09296;%%
\bibitem [{\citenamefont {{Shull}}\ and\ \citenamefont {{van
  Steenberg}}(1985)}]{1985ApJ...298..268S}%
  \BibitemOpen
  \bibfield  {author} {\bibinfo {author} {\bibfnamefont {J.~M.}\ \bibnamefont
  {{Shull}}}\ and\ \bibinfo {author} {\bibfnamefont {M.~E.}\ \bibnamefont {{van
  Steenberg}}},\ }\href {\doibase 10.1086/163605} {\bibfield  {journal}
  {\bibinfo  {journal} {\apj}\ }\textbf {\bibinfo {volume} {298}},\ \bibinfo
  {pages} {268} (\bibinfo {year} {1985})}\BibitemShut {NoStop}%
\bibitem [{\citenamefont {Slatyer}(2016{\natexlab{b}})}]{Slatyer:2015kla}%
  \BibitemOpen
  \bibfield  {author} {\bibinfo {author} {\bibfnamefont {T.~R.}\ \bibnamefont
  {Slatyer}},\ }\href {\doibase 10.1103/PhysRevD.93.023521} {\bibfield
  {journal} {\bibinfo  {journal} {Phys. Rev.}\ }\textbf {\bibinfo {volume}
  {D93}},\ \bibinfo {pages} {023521} (\bibinfo {year} {2016}{\natexlab{b}})},\
  \Eprint {http://arxiv.org/abs/1506.03812} {arXiv:1506.03812 [astro-ph.CO]}
  \BibitemShut {NoStop}%
%%CITATION = ARXIV:1506.03812;%%
\bibitem [{\citenamefont {Taylor}\ and\ \citenamefont
  {Silk}(2003)}]{Taylor:2002zd}%
  \BibitemOpen
  \bibfield  {author} {\bibinfo {author} {\bibfnamefont {J.~E.}\ \bibnamefont
  {Taylor}}\ and\ \bibinfo {author} {\bibfnamefont {J.}~\bibnamefont {Silk}},\
  }\href {\doibase 10.1046/j.1365-8711.2003.06201.x} {\bibfield  {journal}
  {\bibinfo  {journal} {Mon. Not. Roy. Astron. Soc.}\ }\textbf {\bibinfo
  {volume} {339}},\ \bibinfo {pages} {505} (\bibinfo {year} {2003})},\ \Eprint
  {http://arxiv.org/abs/astro-ph/0207299} {arXiv:astro-ph/0207299 [astro-ph]}
  \BibitemShut {NoStop}%
%%CITATION = ASTRO-PH/0207299;%%
\bibitem [{\citenamefont {Schneider}\ \emph {et~al.}(2013)\citenamefont
  {Schneider}, \citenamefont {Smith},\ and\ \citenamefont
  {Reed}}]{Schneider:2013ria}%
  \BibitemOpen
  \bibfield  {author} {\bibinfo {author} {\bibfnamefont {A.}~\bibnamefont
  {Schneider}}, \bibinfo {author} {\bibfnamefont {R.~E.}\ \bibnamefont
  {Smith}}, \ and\ \bibinfo {author} {\bibfnamefont {D.}~\bibnamefont {Reed}},\
  }\href {\doibase 10.1093/mnras/stt829} {\bibfield  {journal} {\bibinfo
  {journal} {Mon. Not. Roy. Astron. Soc.}\ }\textbf {\bibinfo {volume} {433}},\
  \bibinfo {pages} {1573} (\bibinfo {year} {2013})},\ \Eprint
  {http://arxiv.org/abs/1303.0839} {arXiv:1303.0839 [astro-ph.CO]} \BibitemShut
  {NoStop}%
%%CITATION = ARXIV:1303.0839;%%
\bibitem [{\citenamefont {Sheth}\ and\ \citenamefont
  {Tormen}(1999)}]{Sheth:1999mn}%
  \BibitemOpen
  \bibfield  {author} {\bibinfo {author} {\bibfnamefont {R.~K.}\ \bibnamefont
  {Sheth}}\ and\ \bibinfo {author} {\bibfnamefont {G.}~\bibnamefont {Tormen}},\
  }\href {\doibase 10.1046/j.1365-8711.1999.02692.x} {\bibfield  {journal}
  {\bibinfo  {journal} {Mon. Not. Roy. Astron. Soc.}\ }\textbf {\bibinfo
  {volume} {308}},\ \bibinfo {pages} {119} (\bibinfo {year} {1999})},\ \Eprint
  {http://arxiv.org/abs/astro-ph/9901122} {arXiv:astro-ph/9901122 [astro-ph]}
  \BibitemShut {NoStop}%
%%CITATION = ASTRO-PH/9901122;%%
\bibitem [{\citenamefont {{Einasto}}(1965)}]{1965TrAlm...5...87E}%
  \BibitemOpen
  \bibfield  {author} {\bibinfo {author} {\bibfnamefont {J.}~\bibnamefont
  {{Einasto}}},\ }\href@noop {} {\bibfield  {journal} {\bibinfo  {journal}
  {Trudy Astrofizicheskogo Instituta Alma-Ata}\ }\textbf {\bibinfo {volume}
  {5}},\ \bibinfo {pages} {87} (\bibinfo {year} {1965})}\BibitemShut {NoStop}%
\bibitem [{\citenamefont {Klypin}\ \emph {et~al.}(2016)\citenamefont {Klypin},
  \citenamefont {Yepes}, \citenamefont {Gottlober}, \citenamefont {Prada},\
  and\ \citenamefont {Hess}}]{Klypin:2014kpa}%
  \BibitemOpen
  \bibfield  {author} {\bibinfo {author} {\bibfnamefont {A.}~\bibnamefont
  {Klypin}}, \bibinfo {author} {\bibfnamefont {G.}~\bibnamefont {Yepes}},
  \bibinfo {author} {\bibfnamefont {S.}~\bibnamefont {Gottlober}}, \bibinfo
  {author} {\bibfnamefont {F.}~\bibnamefont {Prada}}, \ and\ \bibinfo {author}
  {\bibfnamefont {S.}~\bibnamefont {Hess}},\ }\href {\doibase
  10.1093/mnras/stw248} {\bibfield  {journal} {\bibinfo  {journal} {Mon. Not.
  Roy. Astron. Soc.}\ }\textbf {\bibinfo {volume} {457}},\ \bibinfo {pages}
  {4340} (\bibinfo {year} {2016})},\ \Eprint {http://arxiv.org/abs/1411.4001}
  {arXiv:1411.4001 [astro-ph.CO]} \BibitemShut {NoStop}%
%%CITATION = ARXIV:1411.4001;%%
\bibitem [{\citenamefont {Ali-Haimoud}\ and\ \citenamefont
  {Hirata}(2011)}]{AliHaimoud:2010dx}%
  \BibitemOpen
  \bibfield  {author} {\bibinfo {author} {\bibfnamefont {Y.}~\bibnamefont
  {Ali-Haimoud}}\ and\ \bibinfo {author} {\bibfnamefont {C.~M.}\ \bibnamefont
  {Hirata}},\ }\href {\doibase 10.1103/PhysRevD.83.043513} {\bibfield
  {journal} {\bibinfo  {journal} {Phys. Rev.}\ }\textbf {\bibinfo {volume}
  {D83}},\ \bibinfo {pages} {043513} (\bibinfo {year} {2011})},\ \Eprint
  {http://arxiv.org/abs/1011.3758} {arXiv:1011.3758 [astro-ph.CO]} \BibitemShut
  {NoStop}%
%%CITATION = ARXIV:1011.3758;%%
\bibitem [{\citenamefont {Lewis}\ \emph {et~al.}(2000)\citenamefont {Lewis},
  \citenamefont {Challinor},\ and\ \citenamefont {Lasenby}}]{Lewis:1999bs}%
  \BibitemOpen
  \bibfield  {author} {\bibinfo {author} {\bibfnamefont {A.}~\bibnamefont
  {Lewis}}, \bibinfo {author} {\bibfnamefont {A.}~\bibnamefont {Challinor}}, \
  and\ \bibinfo {author} {\bibfnamefont {A.}~\bibnamefont {Lasenby}},\ }\href
  {\doibase 10.1086/309179} {\bibfield  {journal} {\bibinfo  {journal}
  {Astrophys. J.}\ }\textbf {\bibinfo {volume} {538}},\ \bibinfo {pages} {473}
  (\bibinfo {year} {2000})},\ \Eprint {http://arxiv.org/abs/astro-ph/9911177}
  {arXiv:astro-ph/9911177 [astro-ph]} \BibitemShut {NoStop}%
%%CITATION = ASTRO-PH/9911177;%%
\bibitem [{\citenamefont {Lewis}\ and\ \citenamefont
  {Bridle}(2002)}]{Lewis:2002ah}%
  \BibitemOpen
  \bibfield  {author} {\bibinfo {author} {\bibfnamefont {A.}~\bibnamefont
  {Lewis}}\ and\ \bibinfo {author} {\bibfnamefont {S.}~\bibnamefont {Bridle}},\
  }\href {\doibase 10.1103/PhysRevD.66.103511} {\bibfield  {journal} {\bibinfo
  {journal} {Phys. Rev.}\ }\textbf {\bibinfo {volume} {D66}},\ \bibinfo {pages}
  {103511} (\bibinfo {year} {2002})},\ \Eprint
  {http://arxiv.org/abs/astro-ph/0205436} {arXiv:astro-ph/0205436 [astro-ph]}
  \BibitemShut {NoStop}%
%%CITATION = ASTRO-PH/0205436;%%
\bibitem [{\citenamefont {Lewis}(2013)}]{Lewis:2013hha}%
  \BibitemOpen
  \bibfield  {author} {\bibinfo {author} {\bibfnamefont {A.}~\bibnamefont
  {Lewis}},\ }\href {\doibase 10.1103/PhysRevD.87.103529} {\bibfield  {journal}
  {\bibinfo  {journal} {Phys. Rev.}\ }\textbf {\bibinfo {volume} {D87}},\
  \bibinfo {pages} {103529} (\bibinfo {year} {2013})},\ \Eprint
  {http://arxiv.org/abs/1304.4473} {arXiv:1304.4473 [astro-ph.CO]} \BibitemShut
  {NoStop}%
%%CITATION = ARXIV:1304.4473;%%
\bibitem [{\citenamefont {Li}\ \emph {et~al.}(2018{\natexlab{b}})\citenamefont
  {Li}, \citenamefont {Li}, \citenamefont {Liu}, \citenamefont {Li},\ and\
  \citenamefont {Zhang}}]{Li:2018rwc}%
  \BibitemOpen
  \bibfield  {author} {\bibinfo {author} {\bibfnamefont {H.}~\bibnamefont
  {Li}}, \bibinfo {author} {\bibfnamefont {S.-Y.}\ \bibnamefont {Li}}, \bibinfo
  {author} {\bibfnamefont {Y.}~\bibnamefont {Liu}}, \bibinfo {author}
  {\bibfnamefont {Y.-P.}\ \bibnamefont {Li}}, \ and\ \bibinfo {author}
  {\bibfnamefont {X.}~\bibnamefont {Zhang}},\ }\href {\doibase
  10.1038/s41550-017-0373-0} {\bibfield  {journal} {\bibinfo  {journal} {Nat.
  Astron.}\ }\textbf {\bibinfo {volume} {2}},\ \bibinfo {pages} {104} (\bibinfo
  {year} {2018}{\natexlab{b}})},\ \Eprint {http://arxiv.org/abs/1802.08455}
  {arXiv:1802.08455 [astro-ph.IM]} \BibitemShut {NoStop}%
%%CITATION = ARXIV:1802.08455;%%
\bibitem [{\citenamefont {Hazumi}\ \emph {et~al.}(2019)\citenamefont {Hazumi}
  \emph {et~al.}}]{Hazumi:2019lys}%
  \BibitemOpen
  \bibfield  {author} {\bibinfo {author} {\bibfnamefont {M.}~\bibnamefont
  {Hazumi}} \emph {et~al.},\ }\href {\doibase 10.1007/s10909-019-02150-5}
  {\bibfield  {journal} {\bibinfo  {journal} {J. Low. Temp. Phys.}\ }\textbf
  {\bibinfo {volume} {194}},\ \bibinfo {pages} {443} (\bibinfo {year}
  {2019})}\BibitemShut {NoStop}%
%%CITATION = JLTPA,194,443;%%
\bibitem [{\citenamefont {Abazajian}\ \emph {et~al.}(2016)\citenamefont
  {Abazajian} \emph {et~al.}}]{Abazajian:2016yjj}%
  \BibitemOpen
  \bibfield  {author} {\bibinfo {author} {\bibfnamefont {K.~N.}\ \bibnamefont
  {Abazajian}} \emph {et~al.} (\bibinfo {collaboration} {CMB-S4}),\ }\href@noop
  {} {\  (\bibinfo {year} {2016})},\ \Eprint {http://arxiv.org/abs/1610.02743}
  {arXiv:1610.02743 [astro-ph.CO]} \BibitemShut {NoStop}%
%%CITATION = ARXIV:1610.02743;%%
\bibitem [{\citenamefont {Alvarez}\ \emph {et~al.}(2019)\citenamefont {Alvarez}
  \emph {et~al.}}]{Hanany:2019wrm}%
  \BibitemOpen
  \bibfield  {author} {\bibinfo {author} {\bibfnamefont {M.}~\bibnamefont
  {Alvarez}} \emph {et~al.},\ }\href@noop {} {\  (\bibinfo {year} {2019})},\
  \Eprint {http://arxiv.org/abs/1908.07495} {arXiv:1908.07495 [astro-ph.IM]}
  \BibitemShut {NoStop}%
%%CITATION = ARXIV:1908.07495;%%
\bibitem [{\citenamefont {Calabrese}\ \emph {et~al.}(2017)\citenamefont
  {Calabrese}, \citenamefont {Alonso},\ and\ \citenamefont
  {Dunkley}}]{Calabrese:2016eii}%
  \BibitemOpen
  \bibfield  {author} {\bibinfo {author} {\bibfnamefont {E.}~\bibnamefont
  {Calabrese}}, \bibinfo {author} {\bibfnamefont {D.}~\bibnamefont {Alonso}}, \
  and\ \bibinfo {author} {\bibfnamefont {J.}~\bibnamefont {Dunkley}},\ }\href
  {\doibase 10.1103/PhysRevD.95.063504} {\bibfield  {journal} {\bibinfo
  {journal} {Phys. Rev.}\ }\textbf {\bibinfo {volume} {D95}},\ \bibinfo {pages}
  {063504} (\bibinfo {year} {2017})},\ \Eprint
  {http://arxiv.org/abs/1611.10269} {arXiv:1611.10269 [astro-ph.CO]}
  \BibitemShut {NoStop}%
%%CITATION = ARXIV:1611.10269;%%
\bibitem [{\citenamefont {Di~Valentino}\ \emph {et~al.}(2018)\citenamefont
  {Di~Valentino} \emph {et~al.}}]{DiValentino:2016foa}%
  \BibitemOpen
  \bibfield  {author} {\bibinfo {author} {\bibfnamefont {E.}~\bibnamefont
  {Di~Valentino}} \emph {et~al.} (\bibinfo {collaboration} {CORE}),\ }\href
  {\doibase 10.1088/1475-7516/2018/04/017} {\bibfield  {journal} {\bibinfo
  {journal} {JCAP}\ }\textbf {\bibinfo {volume} {1804}},\ \bibinfo {pages}
  {017} (\bibinfo {year} {2018})},\ \Eprint {http://arxiv.org/abs/1612.00021}
  {arXiv:1612.00021 [astro-ph.CO]} \BibitemShut {NoStop}%
%%CITATION = ARXIV:1612.00021;%%
\bibitem [{\citenamefont {Hamimeche}\ and\ \citenamefont
  {Lewis}(2008)}]{Hamimeche:2008ai}%
  \BibitemOpen
  \bibfield  {author} {\bibinfo {author} {\bibfnamefont {S.}~\bibnamefont
  {Hamimeche}}\ and\ \bibinfo {author} {\bibfnamefont {A.}~\bibnamefont
  {Lewis}},\ }\href {\doibase 10.1103/PhysRevD.77.103013} {\bibfield  {journal}
  {\bibinfo  {journal} {Phys. Rev.}\ }\textbf {\bibinfo {volume} {D77}},\
  \bibinfo {pages} {103013} (\bibinfo {year} {2008})},\ \Eprint
  {http://arxiv.org/abs/0801.0554} {arXiv:0801.0554 [astro-ph]} \BibitemShut
  {NoStop}%
%%CITATION = ARXIV:0801.0554;%%
\bibitem [{\citenamefont {Knox}(1995)}]{Knox:1995dq}%
  \BibitemOpen
  \bibfield  {author} {\bibinfo {author} {\bibfnamefont {L.}~\bibnamefont
  {Knox}},\ }\href {\doibase 10.1103/PhysRevD.52.4307} {\bibfield  {journal}
  {\bibinfo  {journal} {Phys. Rev.}\ }\textbf {\bibinfo {volume} {D52}},\
  \bibinfo {pages} {4307} (\bibinfo {year} {1995})},\ \Eprint
  {http://arxiv.org/abs/astro-ph/9504054} {arXiv:astro-ph/9504054 [astro-ph]}
  \BibitemShut {NoStop}%
%%CITATION = ASTRO-PH/9504054;%%
\bibitem [{\citenamefont {Errard}\ \emph {et~al.}(2016)\citenamefont {Errard},
  \citenamefont {Feeney}, \citenamefont {Peiris},\ and\ \citenamefont
  {Jaffe}}]{Errard:2015cxa}%
  \BibitemOpen
  \bibfield  {author} {\bibinfo {author} {\bibfnamefont {J.}~\bibnamefont
  {Errard}}, \bibinfo {author} {\bibfnamefont {S.~M.}\ \bibnamefont {Feeney}},
  \bibinfo {author} {\bibfnamefont {H.~V.}\ \bibnamefont {Peiris}}, \ and\
  \bibinfo {author} {\bibfnamefont {A.~H.}\ \bibnamefont {Jaffe}},\ }\href
  {\doibase 10.1088/1475-7516/2016/03/052} {\bibfield  {journal} {\bibinfo
  {journal} {JCAP}\ }\textbf {\bibinfo {volume} {1603}},\ \bibinfo {pages}
  {052} (\bibinfo {year} {2016})},\ \Eprint {http://arxiv.org/abs/1509.06770}
  {arXiv:1509.06770 [astro-ph.CO]} \BibitemShut {NoStop}%
%%CITATION = ARXIV:1509.06770;%%
\bibitem [{\citenamefont {Alam}\ \emph {et~al.}(2017)\citenamefont {Alam} \emph
  {et~al.}}]{Alam:2016hwk}%
  \BibitemOpen
  \bibfield  {author} {\bibinfo {author} {\bibfnamefont {S.}~\bibnamefont
  {Alam}} \emph {et~al.} (\bibinfo {collaboration} {BOSS}),\ }\href {\doibase
  10.1093/mnras/stx721} {\bibfield  {journal} {\bibinfo  {journal} {Mon. Not.
  Roy. Astron. Soc.}\ }\textbf {\bibinfo {volume} {470}},\ \bibinfo {pages}
  {2617} (\bibinfo {year} {2017})},\ \Eprint {http://arxiv.org/abs/1607.03155}
  {arXiv:1607.03155 [astro-ph.CO]} \BibitemShut {NoStop}%
%%CITATION = ARXIV:1607.03155;%%
\bibitem [{\citenamefont {Beutler}\ \emph {et~al.}(2011)\citenamefont
  {Beutler}, \citenamefont {Blake}, \citenamefont {Colless}, \citenamefont
  {Jones}, \citenamefont {Staveley-Smith}, \citenamefont {Campbell},
  \citenamefont {Parker}, \citenamefont {Saunders},\ and\ \citenamefont
  {Watson}}]{Beutler:2011hx}%
  \BibitemOpen
  \bibfield  {author} {\bibinfo {author} {\bibfnamefont {F.}~\bibnamefont
  {Beutler}}, \bibinfo {author} {\bibfnamefont {C.}~\bibnamefont {Blake}},
  \bibinfo {author} {\bibfnamefont {M.}~\bibnamefont {Colless}}, \bibinfo
  {author} {\bibfnamefont {D.~H.}\ \bibnamefont {Jones}}, \bibinfo {author}
  {\bibfnamefont {L.}~\bibnamefont {Staveley-Smith}}, \bibinfo {author}
  {\bibfnamefont {L.}~\bibnamefont {Campbell}}, \bibinfo {author}
  {\bibfnamefont {Q.}~\bibnamefont {Parker}}, \bibinfo {author} {\bibfnamefont
  {W.}~\bibnamefont {Saunders}}, \ and\ \bibinfo {author} {\bibfnamefont
  {F.}~\bibnamefont {Watson}},\ }\href {\doibase
  10.1111/j.1365-2966.2011.19250.x} {\bibfield  {journal} {\bibinfo  {journal}
  {Mon. Not. Roy. Astron. Soc.}\ }\textbf {\bibinfo {volume} {416}},\ \bibinfo
  {pages} {3017} (\bibinfo {year} {2011})},\ \Eprint
  {http://arxiv.org/abs/1106.3366} {arXiv:1106.3366 [astro-ph.CO]} \BibitemShut
  {NoStop}%
%%CITATION = ARXIV:1106.3366;%%
\bibitem [{\citenamefont {Ross}\ \emph {et~al.}(2015)\citenamefont {Ross},
  \citenamefont {Samushia}, \citenamefont {Howlett}, \citenamefont {Percival},
  \citenamefont {Burden},\ and\ \citenamefont {Manera}}]{Ross:2014qpa}%
  \BibitemOpen
  \bibfield  {author} {\bibinfo {author} {\bibfnamefont {A.~J.}\ \bibnamefont
  {Ross}}, \bibinfo {author} {\bibfnamefont {L.}~\bibnamefont {Samushia}},
  \bibinfo {author} {\bibfnamefont {C.}~\bibnamefont {Howlett}}, \bibinfo
  {author} {\bibfnamefont {W.~J.}\ \bibnamefont {Percival}}, \bibinfo {author}
  {\bibfnamefont {A.}~\bibnamefont {Burden}}, \ and\ \bibinfo {author}
  {\bibfnamefont {M.}~\bibnamefont {Manera}},\ }\href {\doibase
  10.1093/mnras/stv154} {\bibfield  {journal} {\bibinfo  {journal} {Mon. Not.
  Roy. Astron. Soc.}\ }\textbf {\bibinfo {volume} {449}},\ \bibinfo {pages}
  {835} (\bibinfo {year} {2015})},\ \Eprint {http://arxiv.org/abs/1409.3242}
  {arXiv:1409.3242 [astro-ph.CO]} \BibitemShut {NoStop}%
%%CITATION = ARXIV:1409.3242;%%
\bibitem [{\citenamefont {Ade}\ \emph {et~al.}(2018)\citenamefont {Ade} \emph
  {et~al.}}]{Ade:2018gkx}%
  \BibitemOpen
  \bibfield  {author} {\bibinfo {author} {\bibfnamefont {P.~A.~R.}\
  \bibnamefont {Ade}} \emph {et~al.} (\bibinfo {collaboration} {BICEP2, Keck
  Array}),\ }\href {\doibase 10.1103/PhysRevLett.121.221301} {\bibfield
  {journal} {\bibinfo  {journal} {Phys. Rev. Lett.}\ }\textbf {\bibinfo
  {volume} {121}},\ \bibinfo {pages} {221301} (\bibinfo {year} {2018})},\
  \Eprint {http://arxiv.org/abs/1810.05216} {arXiv:1810.05216 [astro-ph.CO]}
  \BibitemShut {NoStop}%
%%CITATION = ARXIV:1810.05216;%%
\bibitem [{\citenamefont {Finkbeiner}\ \emph {et~al.}(2012)\citenamefont
  {Finkbeiner}, \citenamefont {Galli}, \citenamefont {Lin},\ and\ \citenamefont
  {Slatyer}}]{Finkbeiner:2011dx}%
  \BibitemOpen
  \bibfield  {author} {\bibinfo {author} {\bibfnamefont {D.~P.}\ \bibnamefont
  {Finkbeiner}}, \bibinfo {author} {\bibfnamefont {S.}~\bibnamefont {Galli}},
  \bibinfo {author} {\bibfnamefont {T.}~\bibnamefont {Lin}}, \ and\ \bibinfo
  {author} {\bibfnamefont {T.~R.}\ \bibnamefont {Slatyer}},\ }\href {\doibase
  10.1103/PhysRevD.85.043522} {\bibfield  {journal} {\bibinfo  {journal} {Phys.
  Rev.}\ }\textbf {\bibinfo {volume} {D85}},\ \bibinfo {pages} {043522}
  (\bibinfo {year} {2012})},\ \Eprint {http://arxiv.org/abs/1109.6322}
  {arXiv:1109.6322 [astro-ph.CO]} \BibitemShut {NoStop}%
%%CITATION = ARXIV:1109.6322;%%
\bibitem [{\citenamefont {Contaldi}(2016)}]{Contaldi:2016nys}%
  \BibitemOpen
  \bibfield  {author} {\bibinfo {author} {\bibfnamefont {C.~R.}\ \bibnamefont
  {Contaldi}},\ }\href {\doibase 10.1142/S0218271816400149} {\bibfield
  {journal} {\bibinfo  {journal} {Int. J. Mod. Phys.}\ }\textbf {\bibinfo
  {volume} {D25}},\ \bibinfo {pages} {1640014} (\bibinfo {year}
  {2016})}\BibitemShut {NoStop}%
%%CITATION = IMPAE,D25,1640014;%%
\bibitem [{\citenamefont {Henderson}\ \emph {et~al.}(2016)\citenamefont
  {Henderson} \emph {et~al.}}]{Henderson:2015nzj}%
  \BibitemOpen
  \bibfield  {author} {\bibinfo {author} {\bibfnamefont {S.}~\bibnamefont
  {Henderson}} \emph {et~al.},\ }\href {\doibase 10.1007/s10909-016-1575-z}
  {\bibfield  {journal} {\bibinfo  {journal} {J. Low Temp. Phys.}\ }\textbf
  {\bibinfo {volume} {184}},\ \bibinfo {pages} {772} (\bibinfo {year}
  {2016})},\ \Eprint {http://arxiv.org/abs/1510.02809} {arXiv:1510.02809
  [astro-ph.IM]} \BibitemShut {NoStop}%
\bibitem [{\citenamefont {Aiola}\ \emph {et~al.}(2020)\citenamefont {Aiola}
  \emph {et~al.}}]{Aiola:2020azj}%
  \BibitemOpen
  \bibfield  {author} {\bibinfo {author} {\bibfnamefont {S.}~\bibnamefont
  {Aiola}} \emph {et~al.} (\bibinfo {collaboration} {ACT}),\ }\href@noop {} {\
  (\bibinfo {year} {2020})},\ \Eprint {http://arxiv.org/abs/2007.07288}
  {arXiv:2007.07288 [astro-ph.CO]} \BibitemShut {NoStop}%
\bibitem [{\citenamefont {Li}\ \emph {et~al.}(2019)\citenamefont {Li} \emph
  {et~al.}}]{Li:2017drr}%
  \BibitemOpen
  \bibfield  {author} {\bibinfo {author} {\bibfnamefont {H.}~\bibnamefont {Li}}
  \emph {et~al.},\ }\href {\doibase 10.1093/nsr/nwy019} {\bibfield  {journal}
  {\bibinfo  {journal} {Natl. Sci. Rev.}\ }\textbf {\bibinfo {volume} {6}},\
  \bibinfo {pages} {145} (\bibinfo {year} {2019})},\ \Eprint
  {http://arxiv.org/abs/1710.03047} {arXiv:1710.03047 [astro-ph.CO]}
  \BibitemShut {NoStop}%
%%CITATION = ARXIV:1710.03047;%%
\bibitem [{\citenamefont {Creminelli}\ \emph {et~al.}(2015)\citenamefont
  {Creminelli}, \citenamefont {López~Nacir}, \citenamefont {Simonović},
  \citenamefont {Trevisan},\ and\ \citenamefont
  {Zaldarriaga}}]{Creminelli:2015oda}%
  \BibitemOpen
  \bibfield  {author} {\bibinfo {author} {\bibfnamefont {P.}~\bibnamefont
  {Creminelli}}, \bibinfo {author} {\bibfnamefont {D.~L.}\ \bibnamefont
  {López~Nacir}}, \bibinfo {author} {\bibfnamefont {M.}~\bibnamefont
  {Simonović}}, \bibinfo {author} {\bibfnamefont {G.}~\bibnamefont
  {Trevisan}}, \ and\ \bibinfo {author} {\bibfnamefont {M.}~\bibnamefont
  {Zaldarriaga}},\ }\href {\doibase 10.1088/1475-7516/2015/11/031} {\bibfield
  {journal} {\bibinfo  {journal} {JCAP}\ }\textbf {\bibinfo {volume} {1511}},\
  \bibinfo {pages} {031} (\bibinfo {year} {2015})},\ \Eprint
  {http://arxiv.org/abs/1502.01983} {arXiv:1502.01983 [astro-ph.CO]}
  \BibitemShut {NoStop}%
%%CITATION = ARXIV:1502.01983;%%
\bibitem [{\citenamefont {Benson}\ \emph {et~al.}(2014)\citenamefont {Benson}
  \emph {et~al.}}]{Benson:2014qhw}%
  \BibitemOpen
  \bibfield  {author} {\bibinfo {author} {\bibfnamefont {B.}~\bibnamefont
  {Benson}} \emph {et~al.} (\bibinfo {collaboration} {SPT-3G}),\ }\href
  {\doibase 10.1117/12.2057305} {\bibfield  {journal} {\bibinfo  {journal}
  {Proc. SPIE Int. Soc. Opt. Eng.}\ }\textbf {\bibinfo {volume} {9153}},\
  \bibinfo {pages} {91531P} (\bibinfo {year} {2014})},\ \Eprint
  {http://arxiv.org/abs/1407.2973} {arXiv:1407.2973 [astro-ph.IM]} \BibitemShut
  {NoStop}%
\end{thebibliography}%

\end{document}